\documentclass[aps,pra,floatfix,twocolumn,showpacs,tightenlines,groupedaddress,superscriptaddress,amsmath]{revtex4}

\usepackage{amstext}
\usepackage{bbm}
\usepackage{amssymb}
\usepackage{amsmath}
\usepackage{amsfonts}
\usepackage{mathbbol}
\usepackage{placeins}
\usepackage{dsfont}

\ifx\pdftexversion\undefined
  \usepackage[dvips]{color,graphicx}
\else
  \usepackage[pdftex]{color,graphicx}
\fi
\def	\bse{\begin{subequations}}
\def	\ese{\end{subequations}}
\newcommand{\ha}{\hat{a}}
\newcommand{\hadag}{\hat{a}^\dagger}
\newcommand{\be}{\begin{equation}}
\newcommand{\ee}{\end{equation}}
\newcommand{\ra}{\rangle} 
\newcommand{\la}{\langle}
\newcommand{\ad}{a^{\dagger}}

\makeatletter
\newcommand\xleftrightarrow[2][]{%
  \ext@arrow 9999{\longleftrightarrowfill@}{#1}{#2}}
\newcommand\longleftrightarrowfill@{%
  \arrowfill@\leftarrow\relbar\rightarrow}
\makeatother

\begin{document}

\title{Hybrid Topological Quantum Computation with Majorana Fermions: A Cold Atom Setup }
\author{C. Laflamme}
\address{Institute for Quantum Optics and Quantum Information of the Austrian Academy of Sciences, A-6020 Innsbruck, Austria}
\address{Institute for Theoretical Physics, University of Innsbruck, A-6020 Innsbruck, Austria}
\author{M. A. Baranov}
\address{Institute for Quantum Optics and Quantum Information of the Austrian Academy of Sciences, A-6020 Innsbruck, Austria}
\address{Institute for Theoretical Physics, University of Innsbruck, A-6020 Innsbruck, Austria}
\address{RRC `Kurchatov Institute', Kurchatov Square 1, 123182, Moscow, Russia}
\author{P. Zoller}
\address{Institute for Quantum Optics and Quantum Information of the Austrian Academy of Sciences, A-6020 Innsbruck, Austria}
\address{Institute for Theoretical Physics, University of Innsbruck, A-6020 Innsbruck, Austria}
\author{C. V. Kraus}
\address{Institute for Quantum Optics and Quantum Information of the Austrian Academy of Sciences, A-6020 Innsbruck, Austria}
\address{Institute for Theoretical Physics, University of Innsbruck, A-6020 Innsbruck, Austria}

\date{\today}
\begin{abstract}
In this paper we present a hybrid scheme for topological quantum computation in a system of cold atoms trapped in an atomic lattice. A topological qubit subspace is defined using Majorana fermions which emerge in a network of atomic Kitaev one-dimensional wires. We show how braiding can be efficiently implemented in this setup and propose a direct way to demonstrate the non-Abelian nature of Majorana fermions via a single parity measurement. We then introduce a proposal for the efficient, robust and reversible mapping of the topological qubits to a conventional qubit stored in a single atom. There, well-controlled standard techniques can be used to implement the missing gates required for universal computation. Our setup is complemented with an efficient non-destructive protocol to check for errors in the mapping. 

\end{abstract}


\maketitle

% \tableofcontents

%========================================================================================
%========================================================================================
\section{Introduction}
%========================================================================================
The study of topological states of matter is a new direction in the physics of highly entangled quantum matter. Topologically ordered states are host to interesting and unique properties, in particular, quasi-particle excitations which exhibit fractional or non-abelian quantum statistics. The existence of such quasiparticles opens up the possibility for fundamentally new phenomena beyond those attributed to systems of bosons or fermions. One example of such quasi-particles are Majorana fermions (MFs). In addition to their fundamental interest \cite{Wilczek09,Nayak10}, MFs have also been proposed as the basis of a topological quantum computer. There, the interchange (braiding) of MFs is used to perform fault-taulerant gates \cite{Kitaev03, Nayak08, Pachos12, Beenakker13}. 

Proposals for the realisation and braiding of MFs have been proposed in various solid state systems \cite{Fu08,Moore90, Oreg10, SauLutchyn10, Alicea10, Lutchyn10, Alicea11}, and their observation has been reported in superconductor-semiconductor systems \cite{Mourik12,Deng12,Rokhinson12,Das12}. In parallel to the proposal of these solid-state setups, systems supporting MFs and proposals for the reading and writing of a quantum memory have also been proposed in systems of cold atoms trapped in optical lattices \cite{JiangZoller11, Nascimbene12, Wang13, JiangLukin08}. These systems have the advantage of unprecedented control, in particular the possibility to carry out manipulations on individual sites and links of the lattice grid, as has been demonstrated in the groundbreaking experiments \cite{Simon11, Sherson10, Bakr10}.  Based on this experimental progress, an efficient braiding protocol for MFs in an atomic Kitaev wire network has been proposed \cite{Kraus13}, and it has been demonstrated that this protocol is immune to typical experimental errors. 

However, in both solid state and cold atom realisations, controlled access to the topologically protected degrees of freedom remains a challenge. And while braiding of MFs realises quantum gates in a topologically protected way, braiding alone cannot realise all necessary gates for universal computing  \cite{Ahlbrecht09}. A natural compromise is thus a hybrid system, coupling the topologically protected system to a conventional qubit in order to complete the gate set \cite{Bravyi06, DasSarma05}. Many hybrid systems have been proposed in solid state setups \cite{Hassler10, JiangKane11,Pekker13}, where many current protocols require, for example, measurements and state distillation \cite{Sau10,Bonderson11,Flensberg11,Hyart13}. 

 In the following, we propose a system which exploits the control available in atomic setup to implement (i) braiding of MFs in an atomic Kitaev wire setup, (ii) a way to demonstrate the non-Abelian nature of MFs with a single parity measurement, and (iii) a reversible mapping protocol which maps the topological qubit to a conventional qubit stored in a single atom. This mapping allows for the realisation of the missing gates, which can be implemented via well-controlled, standard techniques as well as for the initialisation and readout of the qubit. Together these pieces realise a hybrid scheme which allows for a proof-of-principle demonstration of the use of the braiding of MFs for universal quantum computation in an experimentally realisable atomic setup. 
 
Our proposal is based on zero energy MFs that emerge as quasi-particles with anyonic statistics in a network of atomic one-dimensional ($1$D) quantum wires coupled to a reservoir of fermionic molecules \cite{JiangZoller11}. The degenerate ground state subspace of this system is used to define a qubit subspace, where braiding of the MFs realises topologically protected gates. Using the possibility for single-site and single-link addressing in atomic systems \cite{Simon11, Sherson10, Bakr10}, this topological ground state subspace can be adiabatically mapped to a conventional qubit system, where the quantum information is stored as the presence or absence of an atom on a single site. Once the topological qubit has been mapped to this conventional qubit there are many standard techniques to implement the missing gates \cite{Briegel00}, for example, collisional gates \cite{Jaksch99}, and using the long range Rydberg interaction \cite{Jaksch00, Muller09, Brion07}. Here we consider the use of Rydberg gates, as there exist experimental setups which are able to implement both the Kitaev wire and carry out these gates \cite{Schausz12}.  Finally, an efficient, non-destructive protocol can be carried out to verify if the mapping protocol has been successfully implemented. 

Our analysis includes an analytical solution of the ideal Kitaev wire system and a numerical analysis of the non-ideal system including effects of imperfect single-site/link addressing. In this analysis we do not consider finite temperature effects or effects of interactions between wires. These effects lift the degeneracy of the ground state subspace; this splitting determines the lifetime of the topological qubit and sets an upper bound on the time allowed for adiabatic manipulation of the Majorana fermions. However, this splitting has been shown to be exponentially smaller than the single-particle excitation energy in the wires \cite{SauSarma11}. This exponential suppression ensures that the time scale on which the protocol is carried out can be chosen such that this ground state splitting can be neglected while still satisfying the condition for adiabaticity.  

This article is organized as follows: In Sec.~2 we begin with a review of the emergence of MFs in the Kitaev quantum wire and we briefly explain one possibility to realize this system in a cold atom setup. In Sec.~3 we present an extended discussion of the braiding protocol introduced in \cite{Kraus13}, adding a detailed discussion on the effect of experimental errors and the consequences of an external harmonic trap. In Sec.~4 we explain how a topologically protected logical qubit space can be defined, and we present the set of qubit gates that can be achieved in this setup via braiding. In Sec.~5 we present a protocol that allows for a robust mapping of the topologically protected qubit to a conventional qubit. This protocol is reversible, and allows for preparation and readout of an arbitrary qubit state. Additionally, we show how the long-range interaction of Rydberg states allows for an efficient check if the mapping protocol has been carried out successfully. Finally, in Sec.~6 we show how Rydberg physics can be exploited further to implement the missing gates for universal quantum computation. The combination of the building blocks presented in Sec.~4--6  provides us with a hybrid system connecting a topological protected and a topologically unprotected conventional qubit system for an efficient implementation of a universal quantum computer. 
%========================================================================================
%========================================================================================
\section{Majorana fermions in the Kitaev Chain} \label{sec:system}
%========================================================================================

In this Section we briefly review theoretically the realisation of MFs in the Kitaev quantum wire and introduce their implementation in an atomic setup. This will provide the basis for braiding in Sec~3, the definition of a topological qubit subspace in Sec~4 and the mapping to the conventional qubit subspace, shown in Sec~5.  Then, we explain how the recent advances in cold atom experiments bring an AMO realization of this model into experimental reach. 

\subsection{Theoretical review of Majorana fermions in a Kitaev wire}
\label{sec:theoretical_model}

\begin{figure}[t!]
\begin{center}
	\includegraphics[width=0.45\textwidth]{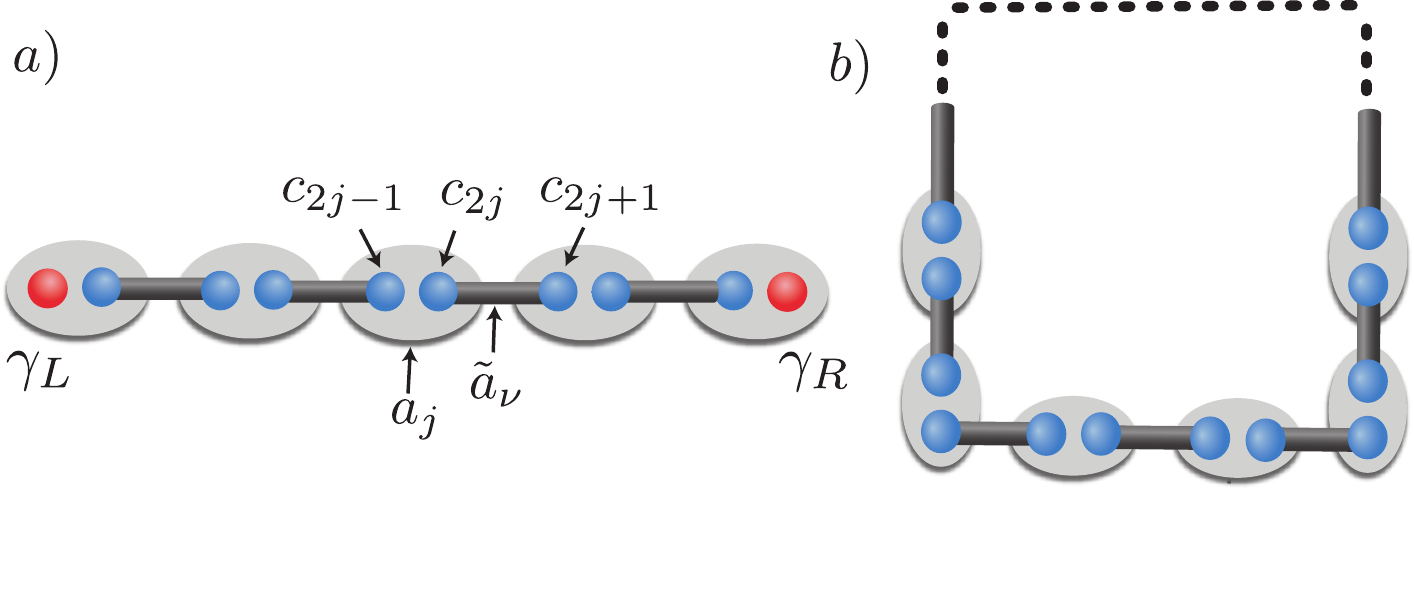}
	\caption{a) A schematic of Kitaev chain with $N$ sites; the $j^{th}$ site corresponds to the fermionic operator $\hat{a}_j$. The Majorana modes $c_{2j-1},c_{2j}$ are defined according to Eq.~(\ref{eq:define_maj_ops}). The fermionic Bogoliubov modes are $\tilde{a}_\nu$, and are composed of Majorana modes from neighbouring sites, leaving two unpaired Majorana modes ($\gamma_{L/R}$) on the ends of the wires. For the ideal chain ($J=\Delta$) these modes are $\gamma_{L}=c_1, \gamma_R=c_{2N}$. For an non-ideal chain ($J\ne \Delta$) these modes decay exponentially into the bulk. \newline b) Enforcing closed boundary conditions links the two end modes forming a closed ring, with no unpaired Majorana modes. }
	\label{fig:define_Kitaev}
\end{center}
\end{figure} 
Following the proposal by Kitaev \cite{Kitaev01}, we consider a system of single component fermions that are confined to a one-dimensional ($1d$)
wire of $N$ sites (see Fig.~\ref{fig:define_Kitaev}a), governed by the Hamiltonian 
\begin{eqnarray}\label{eq:Kitaev}
H_K =\sum_{j=1}^{N-1}H^{(K,J,\Delta)}_{j,j+1} - \mu \sum_j a^{\dagger}_j a_j,
\end{eqnarray}
where we define the coupling Hamiltonian $H^{(K,J,\Delta)}$ between two sites as
\begin{eqnarray}
\label{eq:Kitaev_partII}
H^{(K,J,\Delta)}_{j,j+1}&\equiv& H^{(h,J)}_{j,j+1}+H^{(p,\Delta)}_{j,j+1},
\end{eqnarray}
with
\begin{eqnarray}
H^{(h,J)}_{j,j+1} &\equiv&-J a^{\dagger}_{j}a_{j+1} +{\rm h.c.}, \nonumber \\
H^{(p,\Delta)}_{j,j+1} &\equiv& \Delta a_{j}a_{j+1} +{\rm h.c.} 
\end{eqnarray}
Here, $\ad_j$ and $a_j$, $j = 1\ldots, N$ are fermionic creation and annihilation operators. The parameters $J>0$ and $\Delta \in \mathds{R}$ denote the hopping and pairing amplitudes, respectively, and $\mu$ is a chemical potential.

A Hamiltonian of the form \eqref{eq:Kitaev} can be easily diagonalized in the Majorana representation. There, instead of using the $2N$ creation and annihilation operators, we work with the $2N$ hermitian operators 
\begin{eqnarray}
\label{eq:define_maj_ops}
c_{2j-1}=a_{j}^{\dagger }+a_{j}; \qquad
c_{2j}=(-i)(a_{j}^{\dagger }-a_{j})
\end{eqnarray} that fulfill $\{c_k,c_l\} =  2 \delta_{kl}$. In this representation, the Hamiltonian can be rewritten as $H = i\sum_{k,l}h_{kl}c_kc_l$, where $h = -h^T$ is a real matrix. For the `ideal' quantum wire $(J=|\Delta |,\mu =0)$, the Hamiltonian simplifies to $H =  - iJ\sum_{j=1}^{N-1}c_{2j}c_{2j+1}$. In this case, the Hamiltonian has two zero Majorana modes  $\gamma_{L/R}=c_{1/2N}$ located at the ends of the wire and combine to form a zero-energy non-local fermion $f=\gamma _{L}-i\gamma _{R}$ (Bologoliubov zero-energy mode). The other $N-1$ fermionic Bolgoliubov modes have the form $\tilde{a}_\nu^\dagger=(c_{2\nu+1}+ic_{2\nu}), \nu \in [1,N-1]$ and link neighbouring sites in the bulk (see Fig.~\ref{fig:define_Kitaev}a). In the ideal wire, these modes are degenerate, with energy $2J$.

In the general case $|\Delta| \ne J$, $|\mu| <2J$, the corresponding zero-energy Majorana modes are $\gamma_{L/R} =\sum_j \nu_j^{L/R}c_j$, where the (real) coefficients $\nu_j$ are such that these modes are localised at the left/right edge of the wire, decaying exponentially inside the bulk. The energy of these modes, and of the corresponding non-local fermion $f=\gamma _{L}-i\gamma _{R}$, are not exactly zero (as is the case for the ideal wire) rather the energy scales like $\sim \exp (-N\xi)$ and approaches zero for $N \rightarrow \infty$. The energies of the other $N-1$ fermionic Bogoliubov modes are no longer degenerate, rather they split into an energy band. 

Since the Hamiltonian \eqref{eq:Kitaev} commutes with the fermion-number parity operator $P=(-1)^{\sum_j a^\dagger_ja_j}$, the ground state subspace decomposes into two decoupled sectors. The two degenerate ground states $|+\rangle $ and $|-\rangle $ (with even and odd parity respectively)
correspond to the presence or absence of one non-local fermion $f$: The ground states fulfill the conditions $f|-\rangle =0$ and  $f^{\dagger }|-\rangle =|+\rangle$. 

In the following we call the Kitaev chain of the form in Hamiltonian \eqref{eq:Kitaev} an `open' chain, in reference to the open (unlinked) boundary conditions at each end of the chain (see Fig.~\ref{fig:define_Kitaev} a). Additionally, it will be useful to define a `closed' chain by enforcing periodic boundary conditions which link the last site to the first site, obtaining a closed ring (see Fig.~\ref{fig:define_Kitaev}b). This `closed' chain no longer has a degenerate ground state space; there is one unique (odd parity) ground state $|g\rangle$ and $N$ first excited states. In addition to the $N-1$ excited states shown above given by $\tilde{a}_\nu^\dagger |g\rangle,$ one additional gapped Bogoliubov mode is present due to the periodic boundary conditions. The associated state $\tilde{a}_{N}^\dagger|g\rangle = (c_{1}+ic_{2N})|g\rangle$ is the $N$-th Bogoliubov mode.

\subsection{Implementation of the Kitaev chain with cold atoms}
\label{sec:coldatom_implementation}
The physical implementation of the Kitaev wire has been discussed mainly in a solid state context. In these systems a semi-conductor wire can be coupled to an s-wave superconductor, giving rise to the pairing term necessary in the Kitaev Hamiltonian \cite{Lutchyn10, Oreg10}. Here, we pursue a complementary route considering a system of cold atoms confined to an optical lattice. 

\begin{figure}[t!]
	\begin{center}
		\includegraphics[width=0.4\textwidth]{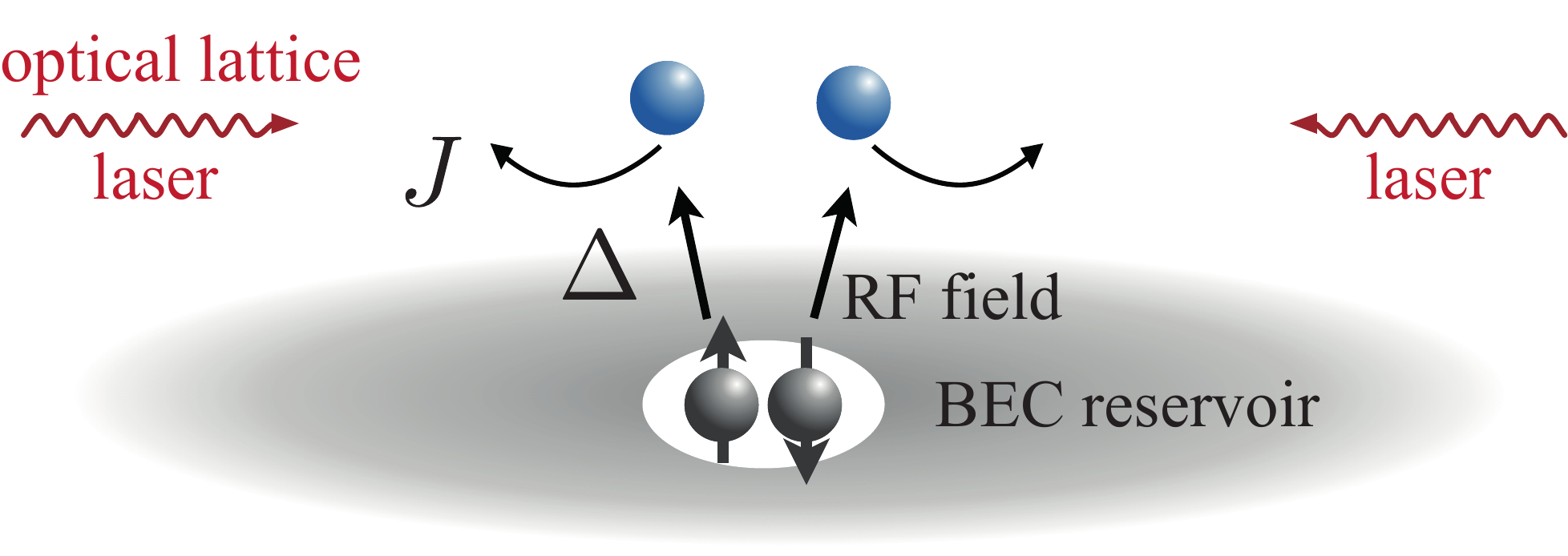}
		\caption{The experimental realisation of a Kitaev wire in an optical lattice setup. The hopping term $\sim J$ comes naturally in an optical lattice setup, while the pairing $\sim \Delta$ is realised by the dissociatation of a molecule from the BEC reservoir with an RF field. A Raman coupling creates an effective spin-orbit coupling as well as a magnetic field giving an effective single component model \cite{JiangZoller11}.}
		\label{fig:BEC_setup}
	\end{center}
	\end{figure}

In the following, we briefly summarise the atomic realization of the Kitaev wire proposed in \cite{JiangZoller11}. We assume a tunnelling, tight-binding model where the hopping term of the Hamiltonian in Eq.~\eqref{eq:Kitaev} ($\ad_j a_{j+1} + {\rm h.c.}$) arises naturally. In Ref.~\cite{JiangZoller11} it has been shown that the pairing term, $\ad_j \ad_{j+1} + {\rm h.c.}$, can be engineered via the coupling of the system to a BEC reservoir of Feshbach molecules. The main idea is to couple the two internal spin states $(\ad_{p, \uparrow}, \ad_{p, \downarrow})$ of the trapped atoms with momentum $p$ in a $1$D wire of  fermions to a Feshbach molecule via an RF field. This results in an effective pairing term of the form $\Delta \ad_{p, \uparrow}\ad_{-p, \downarrow}+h.c$, as shown schematically in Fig.~\ref{fig:BEC_setup}. Additional lasers generate optical Raman transitions with photon recoil to create both an effective magnetic field and an effective spin-orbit coupling, which project out one of the spin components, such that we obtain the spinless pairing term of Eq. \eqref{eq:Kitaev} ~\cite{JiangZoller11}.  Estimates give an energy gap separating the Majorana states from the rest of the spectrum on the order of tens of nano Kelvins. 
For an alternative proposal see Ref. \cite{Nascimbene12}. 
	
Our approach is motivated by several groundbreaking experiments which have proven that the control available in atomic systems provides a unique platform for studying quantum states. This includes the possibility for optical lattices to be loaded in a well-controlled way, as shown by \cite{Greiner02}. This experiment was followed by a breakthrough in controllability, allowing for atom addressing and imaging at the level of single sites \cite{Bakr10,Sherson10}. This possibility for single site addressing is integral for the implementation of braiding shown in Sec~3 and the mapping to the conventional qubit subspace shown in Sec~5. 

%%%%%%%%%%%%%%%%%%%%%%%%%%%%%%%%%%%%%%%%%%%%%%%%%%%%%%%%%%%%%%%%%%	
\section{Braiding of Majorana Fermions}
\label{sec:braiding}
\begin{figure*}[t!]
	\begin{center}
		\includegraphics[width=0.95\textwidth]{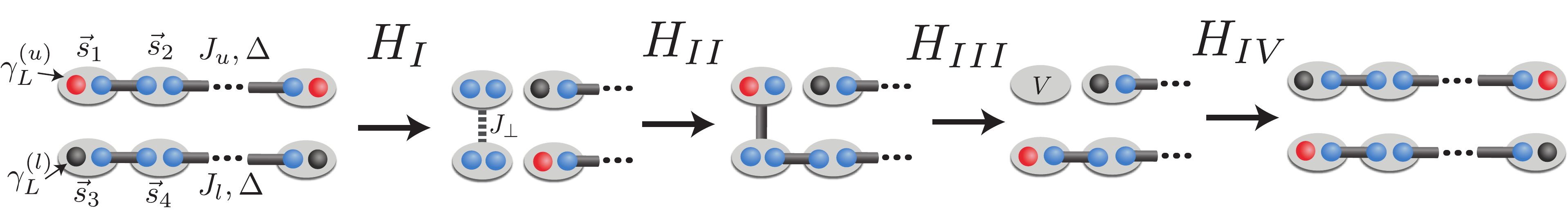}
		\caption{Four step braiding protocol for two ideal quantum wires. The zero-energy Majorana modes $\gamma_L^{(u)},\gamma_L^{(l)}$ that are initially on the upper and lower chains are shown as red and black, respectively. Majorana modes which are blue correspond to those coupled into finite-energy fermionic modes. The coupling of sites via hopping on the upper (lower) chain ($J_{u(l)}$) and pairing ($\Delta$) (Kitaev coupling) is indicated by grey solid links, while the coupling via hopping only ($J_\perp$) is shown as a square-dashed link.}
		\label{fig:braid_steps}
	\end{center}
	\end{figure*}
The braiding, or interchange, of two Majorana modes $\gamma_1$ and $\gamma_2$ gives rise to the non-trivial transformation $\gamma_{1}\mapsto-\gamma_{2}$, $\gamma_{2}\mapsto\gamma_{1}$. This remarkable property can be used as the basis to form topologically protected gates \cite{Nayak08, Kitaev03, Pachos12}.  In this section we present a protocol on how to efficiently braid atomic MFs which can be realised in an optical lattice implementation of a set of Kitaev wires. We present an extended analysis based on the protocol introduced in \cite{Kraus13}; here we include an extended discussion on possible experimental errors, including the effects of an external harmonic trap, as well as including a proposal for the direct demonstration of the non-Abelian statistics of MFs via a single parity measurement.

We consider two Kitaev wires that are aligned in parallel as depicted in  Fig.~\ref{fig:braid_steps}. In section~\ref{sec:ideal_braiding}  we consider the case of ideal wires, which we solve analytically. In section~\ref{sec:imperfect_braiding} we solve numerically the experimentally realistic scenario of non-ideal wires. Additionally we will further consider the effect of an external harmonic trapping potential. We label the sites on the two wires by $(w,j)$, where $w=u,l $ denotes the upper resp. lower wire and $j=1,\ldots ,N$ enumerate the sites in the one-dimensional configuration.  Each wire is described by a Hamiltonian $H^{(w)}$ of the form given in Eq.~\eqref{eq:Kitaev}, with $j \rightarrow (w,j)$, $J, \Delta, \mu \mapsto J_w, \Delta_w, \mu_w$. In the following, only operations on the two sites  $(u,1)$ and $(l,1)$ on the left side of the wire  and the nearby links are required. To simplify notation, we label the involved sites by  $\vec{s}_{1}=(u,1)$, $\vec{s}_{2}=(u,2)$,$\vec{s}_{3}=(l,1)$ and $\vec{s}_{4}=(l,2)$ (see Fig.~\ref{fig:braid_steps}).  %Further, we write $\hat u_{j}\equiv c_{u,j}$ for the Majorana operators on the upper wire  and $\hat l_{j} \equiv c_{l,j}$ for the Majorana operators on the lower wire. 
We start with an analysis of two ideal wires, i.e. $J_u = \Delta_u >0, J_l = \Delta_l>0, \mu_{u,l} = 0$, as this case allows for a simple analytic treatment. We assume w.l.o.g. $|J_u|>J_l$. Extending the analysis to non-ideal wires will be done numerically in the following section. 
\subsection{Ideal case}
\label{sec:ideal_braiding} 
	
In the case of two ideal wires, the Majorana modes on the upper and lower wire are of the form $\gamma _{L}^{(u)}=\hat c_{u,1}$, $\gamma _{R}^{(u)}=c_{u,2N}$, $\gamma_{L}^{(l)}= c_{l,1}$, $\gamma _{R}^{(l)}= c_{l,2N}$. We introduce now a protocol that allows for the braiding of the left Majorana modes $\gamma _{L}^{(u)}$ and 
$\gamma _{L}^{(l)}$:
\begin{align}
\label{eq:braiding_twowires}
\gamma _{L}^{(u)} &\mapsto \gamma _{L}^{(l)} \nonumber \\
\gamma_{L}^{(l)} &\mapsto -\gamma _{L}^{(u)}.
\end{align}
This braiding is done by adiabatically changing the Hamiltonian on the left side of the wire in four steps, as shown in Fig.~\ref{fig:braid_steps}. The protocol requires the ability to switch on/off (i) the hopping $H_{\vec{s}_{i},\vec{s}_{j}}^{(h,J)}=-Ja_{%
\vec{s}_{i}}^{\dagger }a_{\vec{s}_{j}}+{\rm h.c.}$ and (ii) the pairing $H_{\vec{s}%
_{i},\vec{s}_{j}}^{(p,\Delta)}=\Delta a_{\vec{s}_{i}}a_{\vec{s}_{j}}+{\rm h.c.}$ between the
neighboring sites $\vec{s}_{i}$ and $\vec{s}_{j}$ and (iii) the local
potential $H_{\vec{s}_{i}}^{({V)}}=2Va_{\vec{s}_{i}}^{\dagger }a_{\vec{s}%
_{i}}$ on site $\vec{s}_{i}$. Note that a combination of (i) and (ii) allows
to switch on/off the (Kitaev) coupling $H_{\vec{s}_{i},\vec{s}_{j}}^{(K, J, \Delta)}=H_{%
\vec{s}_{i},\vec{s}_{j}}^{(h,J)}+H_{\vec{s}_{i},\vec{s}_{j}}^{(p,\Delta)} \equiv H_{\vec{s}_{i},\vec{s}_{j}}^{(K, J)}$, since $J = \Delta$ in this case. These
operations rely on the possibility to address single sites or links in cold
atom experiments, as demonstrated in \cite{Bakr10,Sherson10}. 

We describe now in detail the four steps of the braiding protocol. The underlying physical process
of the braiding protocol is the transfer of one fermion from the system (i. e. either from the
upper or from the lower wire) into the lower wire. To this end, we decouple first the sites $\vec s_1$ and $\vec s_2$ from the rest of the chain, such that when they are completely decoupled they carry one fermion that has been extracted from the original system. Then, we transfer this fermion to the lower (or upper) wire.  Finally, we restore the original configuration. In the following, we parametrize the adiabatic changes in each step of duration $t_f$ via two continuous and monotonic time-dependent functions $C_t,S_t: [0, J\,t_{f}] \rightarrow [0,1]$, with $C(0) = 1, C(t_{f}) = 0$ and $S(0) = 0, S(t_{f})=1$. To simplify the presentation, we will only write down the Hamiltonian for the four involved
sites in each step.  We follow the evolution of the zero modes which are always separated
by a finite gap $\Delta E $ from the rest of the spectrum. 

\begin{figure*}[t!]
	\begin{center}
		\includegraphics[width=0.95\textwidth]{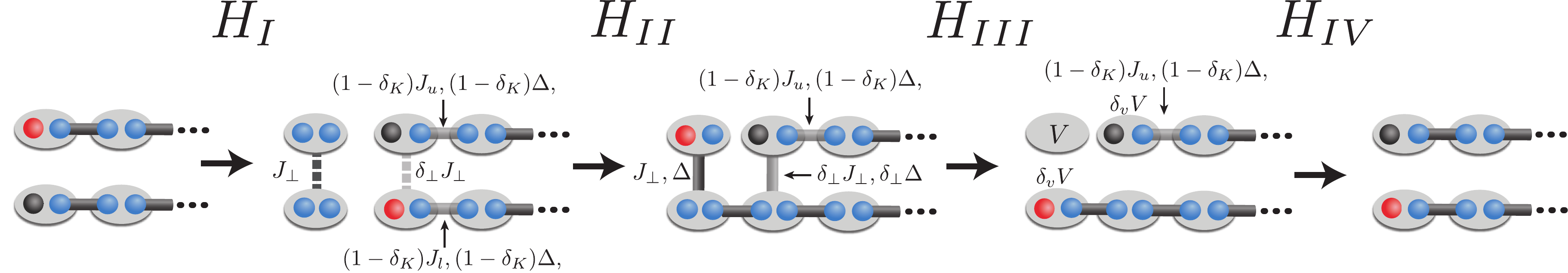}
		\caption{Errors occurring at each step in the braiding protocol. The errors include: An non-ideal chain $J\ne \Delta$, differing couplings on each chain $J_u \ne J_l$, and cross talk due to imperfections in local operations which implies switching on couplings between two sites induces a coupling between adjacent sites with strength $0<\delta_k, \delta_\perp<1$ and turning on a potential on a site $i$ induces a potential on neighbouring sites with strength $0<\delta_v<1$. }
		\label{fig:braiding_steps_errors}
	\end{center}
	\end{figure*}
	
In the first step we decouple the two left most sites, $\vec{s}_{1}$ and $\vec{s%
}_{3}$ from the system by switching off the couplings $H_{\vec{s}_{i},\vec{s}%
_{j}}^{(K)}$ between sites $\vec{s}_{1}-\vec{s}_{2}$ and $\vec{s}_{3}-\vec{s}%
_{4}$, and, at the same time, switching on a hopping of strength $J_\perp$ between sites $\vec{s}%
_{1}-\vec{s}_{3}$: 
\begin{align*}
& H_{I}(t)= C_t\left[H_{\vec{s}_{1}\vec{s}_{2}}^{(K, J_u)}+H_{\vec{s}_{3}%
	\vec{s}_{4}}^{(K, J_l)}\right]+S_tH_{\vec{s}_{1}\vec{s}_{3}}^{(h, J_\perp)} 
\end{align*}%
By solving the Heisenberg equations of motion, we find the evolution of the zero modes satisfy
\begin{align*}
\gamma_{L}^{(u)}(t) &= \frac{2 J_l C_tc_{u,1}-J_\perp S_t c_{l,3}}{\sqrt{4 J_l^2 C_t^2 + J_\perp^2 S_t^2}},\\
\gamma_{L}^{(l)}(t) &= \frac{2 J_u C_tc_{l,1}-J_\perp S_t c_{u,3}}{\sqrt{4 J_u^2 C_t^2 + J_\perp^2 S_t^2}},
\end{align*}
such that at $t=t_f$, $\gamma _{L}^{(u)}(t_f)=-c_{l,3}$ and $\gamma _{L}^{(d)}(t_f)=-c_{u,3}$. These zero modes are always separated by the gap $\Delta E(t) =\sqrt{J_{l}^2 C_t^2 + J_\perp^2 S_t^2}$ from the rest of the spectrum. At the end of Step I the two sites $\vec{s}_{1}$ and $\vec{s}_{3}$  are independent of the rest of the system and are coupled to one another with hopping parameter $J_\perp$. As the adiabatic theorem ensures that the system remains in the ground state throughout the entire evolution, at the end of Step I, exactly one fermion will occupy the symmetric superposition state on these two sites. As we will discuss in the following section, this extraction of one particle from the system will contribute to the robustness of the protocol against errors. \newline

%%%%%%%%
In the second step we put now this fermion in the lower wire by switching on $%
H_{\vec{s}_{i},\vec{s}_{j}}^{(K, J_l)}$ between sites $\vec{s}_{3}-\vec{s}_{4}$,
and $H_{\vec{s}_{i},\vec{s}_{j}}^{(p, J_\perp)}$ between the sites $\vec{s}_{1}-\vec{s%
}_{3}$: 
\begin{align*}
& H_{II}(t)=H_{\vec{s}_{1}\vec{s}_{3}}^{(h, J_\perp)}+S_t\left[ H_{\vec{s}%
_{1}\vec{s}_{3}}^{(p, J_\perp)}+H_{\vec{s}_{3}\vec{s}_{4}}^{(K, J_l)}\right].
\end{align*}%
The zero modes evolve as 
\begin{align*}
\gamma _{L}^{(u)}(t)&=\frac{2J_lS_tc_{u,1}-J_\perp (1-S_t)c_{l,3}}{\sqrt{4 J_l^2 S_t^2 + J_\perp^2 (1-S_t)^2}},\\
\gamma _{L}^{(l)}(t)&=-c_{u,3},
\end{align*}
 such that at
the end $\gamma _{L}^{(u)}(t_f)=c_{u,1}$ and $\gamma _{L}^{(d)}(t_f)=-c_{u,3}$. The gap is given by $\Delta E (t)= \min (\Delta E_1, \Delta E_2) > 0$, where $\Delta E_1 = J_\perp (1+ S_t)$ and $\Delta E_2 = \sqrt{J_\perp^2 (1-S_t)^2 + 4J_l^2 S_t^2}$. Note, that
at this stage the Majorana mode $\gamma _{L}^{(u)}$ ($\gamma _{L}^{(l)}$)
has already been moved from the upper (lower) to the lower (upper) wire.
However, two additional steps are needed to recover the original
configuration of the wires. \newline
In the third step we move the Majorana mode from the site $\vec{s}_{1}$ to
the site $\vec{s}_{3}$ by switching on the local potential $H_{\vec{s}_{1}}^{(V)}$ and
simultaneously switching off the coupling $H_{\vec{s}_{i}\vec{s}_{j}}^{(K, J_\perp)}$ between the
sites $\vec{s}_{1}-\vec{s}_{3}$: 
\begin{align*}
H_{III}(t)& =S_tH_{\vec{s}_{1}}^{(V)}+C_tH_{\vec{s}%
_{1}\vec{s}_{3}}^{(K, J_\perp)}+H_{\vec{s}_{3}\vec{s}_{4}}^{(K, J_l)}\end{align*}%
The evolution of the zero mode
\begin{align*}
\gamma _{L}^{(u)}(t)&=\frac{J_\perp C_tc_{u,1}+VS_tc_{l,1}}{\sqrt{J_\perp^2 C_t^2 +V^2 S_t^2}},
\end{align*}
results in $\gamma _{L}^{(u)}(t_f)=c_{l,1}$, while $\gamma_{L}^{(l)}(t)=-c_{u,3}$ remains fixed. The energy gap is given by $\Delta E(t) = \min (J_l, 2\sqrt{J_\perp^2 C_t^2 + V^2 S_t^2})$.\newline
In the fourth and final step we switch off the local potential $H_{\vec{s}_{1}}^{(V)}$ and switch
on the coupling $H_{\vec{s}_{1}\vec{s}_{2}}^{(K, J_u)}$ between sites $\vec{s}_{1}-\vec{s}_{2}$: 
\begin{align*}
H_{IV}(t)& =S_tH_{\vec{s}_{1}\vec{s}_{2}}^{(K, J_u)}+H_{\vec{s}_{3}%
\vec{s}_{4}}^{(K, J_l)}+C_tH_{\vec{s}_{1}}^{(V)} \end{align*}%
The energy gap is calculated to be $\Delta E(t) = \min (J_l, 2\sqrt{V^2 C_t^2 + J_\perp^2 S_t^2})$  and  the zero modes are given by $\gamma _{L}^{(u)}=c_{u,3}$ and
\begin{align}
 \gamma_{L}^{(l)}(t) =- \frac{JS_tc_{u,1}+VC_tc_{u,3}}{\sqrt{(JS_t)^{2}+(VC_t)^{2}}}.
\end{align}
Thus, steps I-IV lead to the desired braiding of the Majorana modes in the left edge of the two wires; corresponding to, up to a trivial phase factor, the unitary 
\begin{equation}
\label{eq:braiding_unitary}
U_{ul}=e^{\pi \gamma _{L}^{(u)}\gamma _{L}^{(l)}/4}.
\end{equation}
 Note that the braiding in the other direction, $U_{ul}^{\dagger }$ and $%
\gamma _{L}^{(u)}\mapsto -\gamma _{L}^{(l)}$, $\gamma _{L}^{(l)}\mapsto
\gamma _{L}^{(u)}$, can be achieved by putting the uncoupled fermion in the
upper (instead of the lower) wire with a simple modification of Steps II-IV.

\begin{figure}[tbp]
\includegraphics[width=0.95\columnwidth]{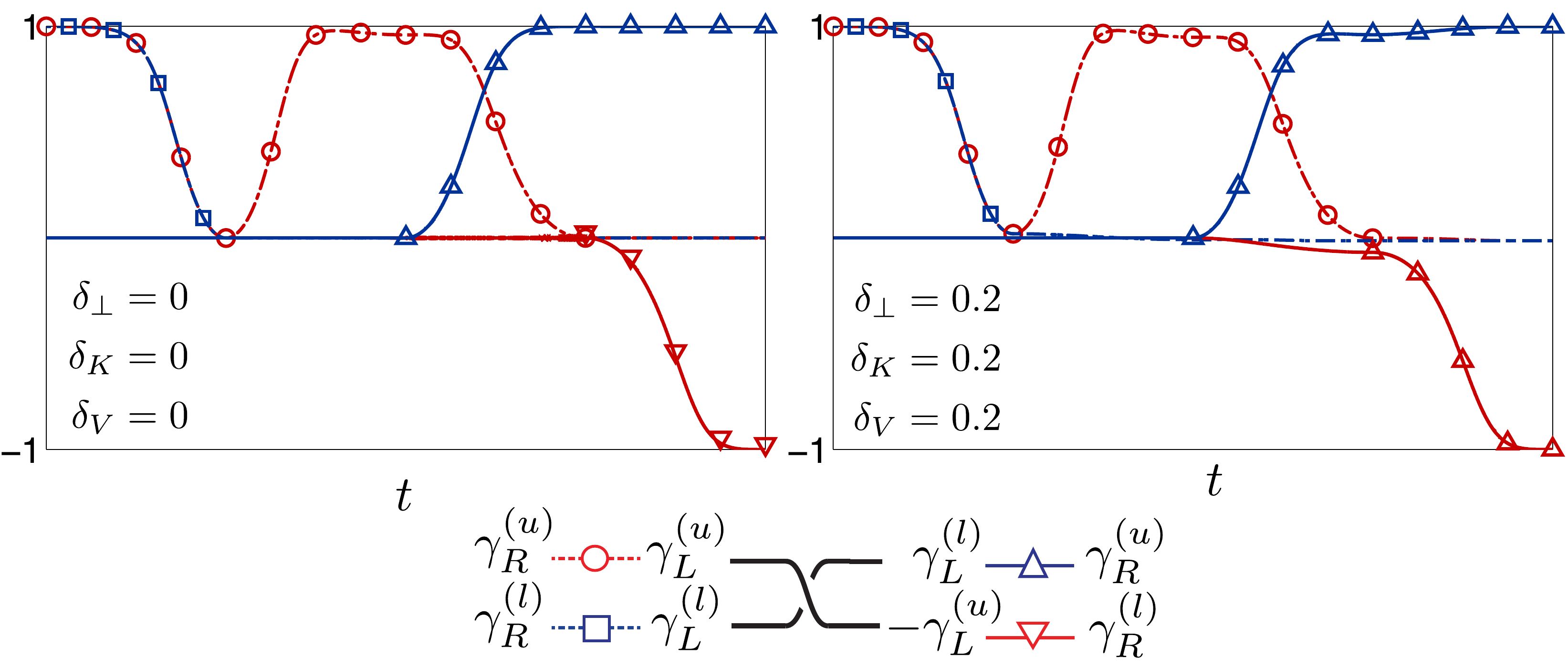}
\caption{Evolution of the Majorana correlation functions $\langle i\protect%
\gamma _{L}^{(u)}\protect\gamma _{R}^{(u)}\rangle $ (red, $\circ$), $\langle i%
\protect\gamma _{L}^{(l)}\protect\gamma _{R}^{(l)}\rangle $ (blue, $\square$), $%
\langle i\protect\gamma _{L}^{(l)}\protect\gamma _{R}^{(u)}\rangle $ (blue,  $\bigtriangleup$), and $\langle i\protect\gamma _{L}^{(u)}\protect\gamma %
_{R}^{(l)}\rangle $ (red, $\bigtriangleup$) during the braiding protocol with errors $%
\delta_K,\delta_v,\delta_\perp $ in the local operations for two non-ideal quantum wires
with $|\Delta |=1.5J$ and $\protect\mu =0$. Markers are only drawn in regions where the correlation functions are non-zero. }
\label{fig:nonperfect}
\end{figure}

The braiding results in the change of the correlation functions of the
Majorana operators (see Fig.~\ref{fig:nonperfect}) and thus also results in the change of the long-range
fermionic correlations. This can also be translated into the change of the
fermionic parities of the wires: If $|+_{w}\rangle $ ($|-_{w}\rangle $)
denotes the state of the $w=u,l$ wire with even (odd) parity and, for
example, we start from the state $|+_{u}+_{l}\rangle $ with both wires
having even parity, then the braiding $U_{ul}$ results in $U_{ul}|+_{u}+_{l}%
\rangle =(|+_{u}+_{l}\rangle +|-_{u}-_{l}\rangle )/\sqrt{2}$, and $%
U_{ul}^{2}|+_{u}+_{l}\rangle =|-_{u}-_{l}\rangle $. The result of the
braiding, therefore, can be checked by measuring the change of the Majorana
correlation functions in Time-of-Flight or spectroscopic experiments \cite{Kraus12}, or
by measuring the parity of the wires by counting the number of fermions
modulo two \cite{Sherson10}.

%%%%%%%%%%%%%%%%%%%%%%%%%%%%%%%%%%%%%%%%%%%%%%%%

\subsection{Effects of imperfections in a cold atom setup}
\label{sec:imperfect_braiding}	

We have just demonstrated
the braiding for the case of ideal Kitaev wires and perfect local operations
(single site/link addressing). Since none of these assumptions are experimentally realistic, we now present a detailed discussion of the effect of relevant  experimental errors on the braiding protocol. 

\subsubsection{Non-Ideal Wires and Imperfect Addressing}
\begin{figure}[tbp]
\includegraphics[width=0.95\columnwidth]{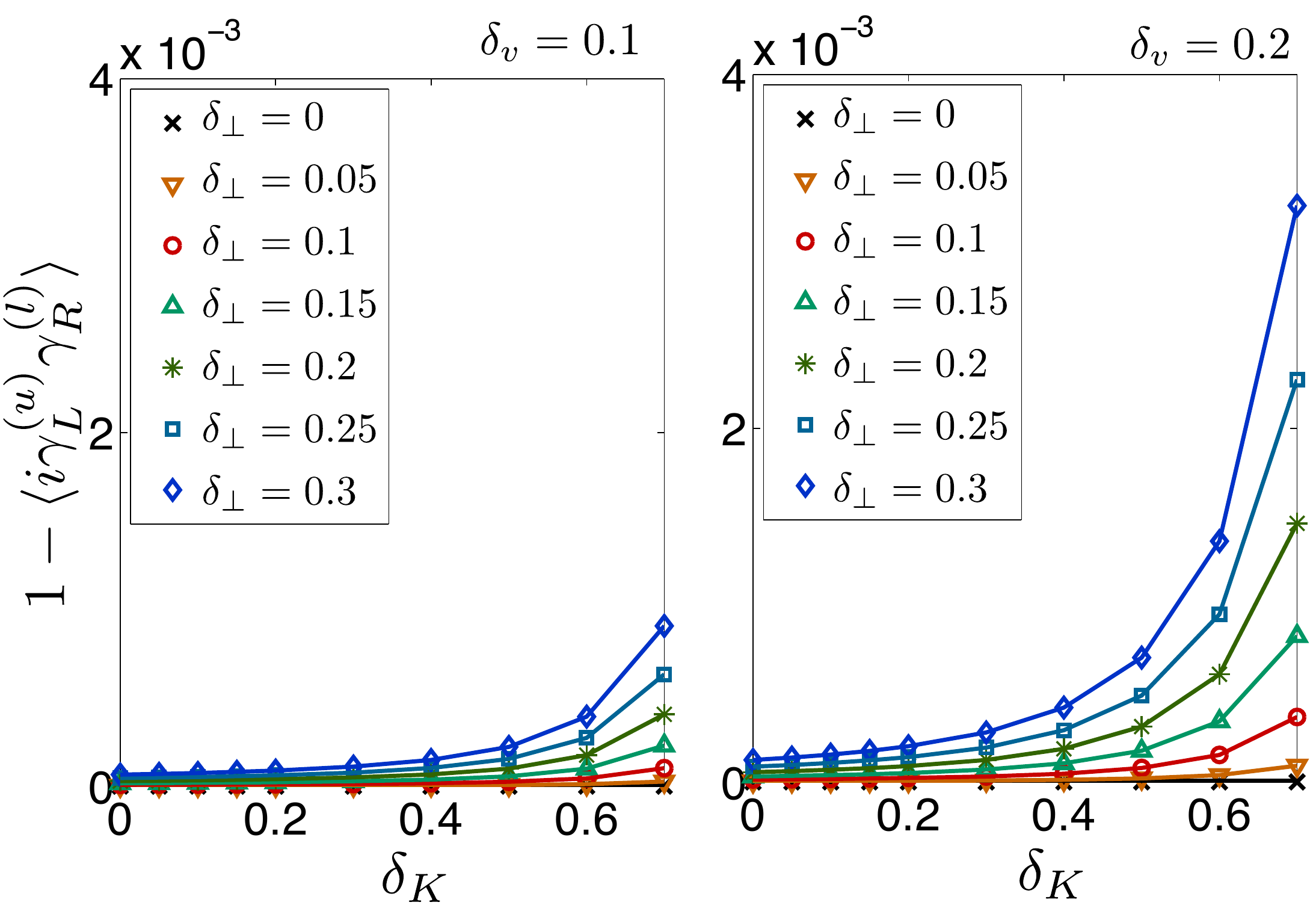}
\caption{The error of the braiding protocol, defined as $\langle i\protect\gamma _{L}^{(l)}\protect\gamma _{R}^{(u)}\rangle $ for increasing error $\delta_K$. The results for $\delta_\perp= 0,0.05,0.1,0.15,0.2,0.25,0.3$ are shown. In the left panel $\delta_V=0.1$ and in the right panel $\delta_V=0.2$. The origin of each error is shown explicitly in Fig.~\ref{fig:braiding_steps_errors}.}
\label{fig:errors}
\end{figure}

First, we relax the assumption of the ideal wire, $J\neq |\Delta |$, $\mu \neq 0$. Then, we assume cross talk induced by imperfect single-site and single-link addressing. This implies the following: (i) Switching on the hopping $J$ and/or the pairing $\Delta $ between the sites $(u,1)-(l,1)$ on the upper and lower wire, also introduces the hopping $J\delta_\perp$ and/or the pairing $\delta_\perp \Delta $ between the adjacent sites $(u,2)-(l,2)$, where $0 \leq \delta_\perp  \leq 1$. (ii) Switching off the couplings between the sites $(w,1)-(w,2)$ also reduces the couplings
between the sites $(w,2)-(w,3)$ by a factor $(1-\delta_K)$, where $0 \leq \delta_K \leq 1$. (iii) Raising the
local potential $V$ on the site $(u,1)$ results in a local potential $\delta_v V$ ($0 \leq \delta_v \leq 1$) on the neighboring sites $(u,2)$ and $(l,1)$. These errors are shown as they appear in each step of the braiding protocol in Fig.~\ref{fig:braiding_steps_errors}. Since it is not possible to give a simple analytic solution for the braiding protocol for these cases, we have carried out a numerical analysis for a system of $N=40$ sites with $\Delta = 1.5 J$. Current experimental techniques have an error in the single-site addressing of about $10\%$. It is a reasonable assumption to take errors in the addressing on an individual wire larger than the errors in operations that involve both wires due to a difference in the wave function overlap. Thus, in the following, we present numerical results for $0\leq \delta_\perp \leq 0.3$, $0 \leq \delta_K \leq 0.7$ and $0 \leq \delta_v \leq 0.2$. The results are given in Fig.~\ref{fig:errors} for $\delta_v = 0.1, 0.2$. For $\delta_v = 0$ the error is less than $10^{-4}$  for all $\delta_\perp, \delta_K$ in the given parameter regime. As can be concluded from the Figure, the braiding protocol is remarkable robust against relatively strong errors. We also conclude the system is more robust against errors on operations on an individual wire than against those involving operations that couple both wires, since the latter lead to a coupling of the Majorana fermions.

In addition to the experimental errors listed above, the protocol described above assumed adiabatic evolution. In general the adiabaticity criteria would imply that $T_f$, the total time of the protocol, should be much larger than the inverse gap. Here, this implies $T_f \gg \hbar/J$. However, numerical simulations show that already for $T_f = \pi/2 \hbar/J$, we obtain a fidelity which deviates from unity with $\sim 10^{-6}$. For hopping of the order of $J \sim 500$Hz, this corresponds to a total time duration of the protocol of the order of milliseconds.

\subsubsection{Influence of an external harmonic confinement}
In the previous Section we have discussed the effects of experimental errors on the braiding protocol that stem from imperfect operations and non-ideal wires. Many experiments with cold atoms have an external trapping potential that might also have harmful effects on the braiding protocol. In the following we discuss the influence of a harmonic external confining potential. Typically, such a confining potential is shallow, i.e. the
potential changes only slightly on the period of the lattice. We show below
that the presence of such a potential does not influence the results of the
braiding. We model the harmonic potential via $V_{\mathrm{trap}}(x_{j}%
)=V_{t}((L+1)/2-j)^{2}/L^{2}$, where $x_{j}=ja$ is the position of lattice
site $j=1,\ldots L$, $a$ is the spacing, and $V_{t}$ measures the strength.
This potential is added to the potential from the lasers configuration which
creates the finite lattice with $L$ sites.

\begin{figure}[t]
\begin{center}
\includegraphics[width=0.95\columnwidth]{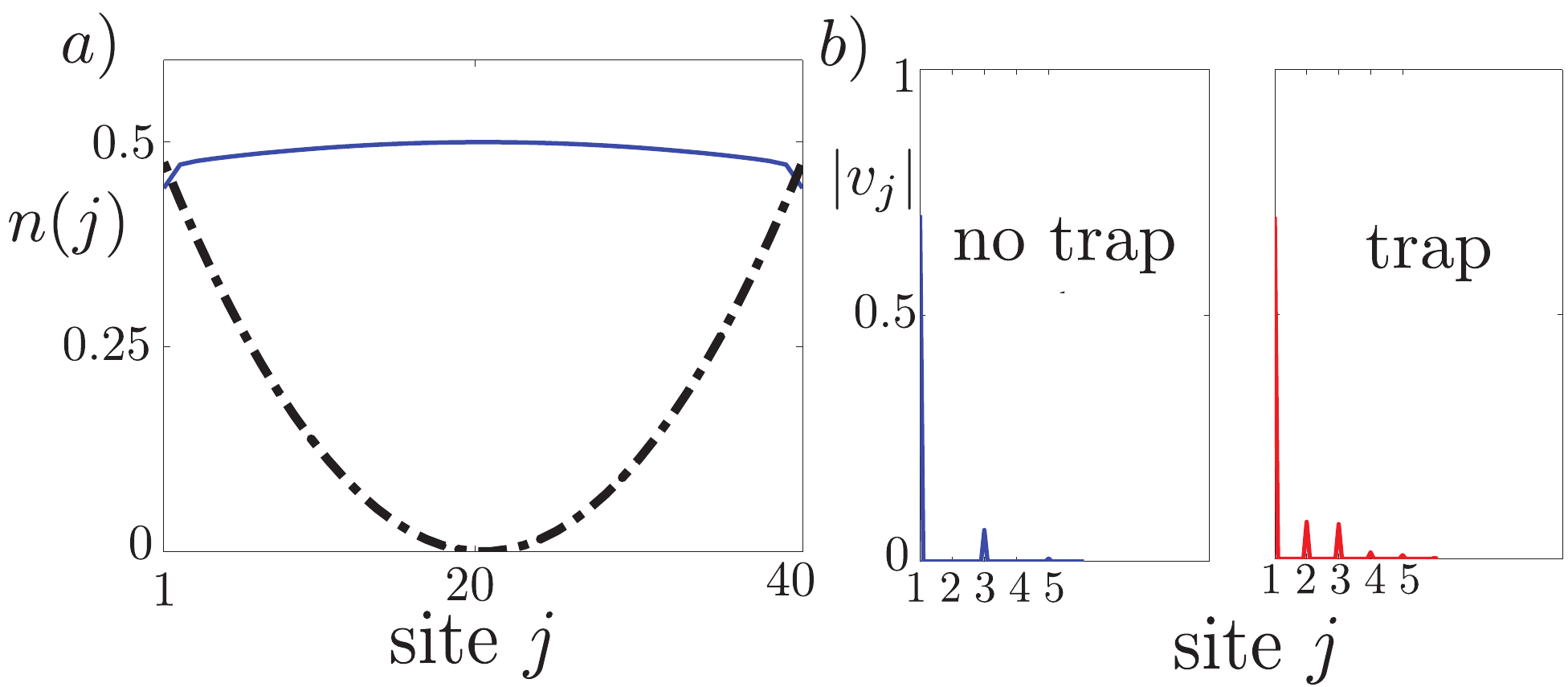}
\end{center}
\caption{Influence of a shallow harmonic trap on a Kitaev chain (see text): a)
The density distribution, b) Physical extent of the Majorana modes.}%
\label{fig:harmonic_density_modes}
\end{figure}

\begin{figure}[t]
\begin{center}
\includegraphics[width=0.75\columnwidth]{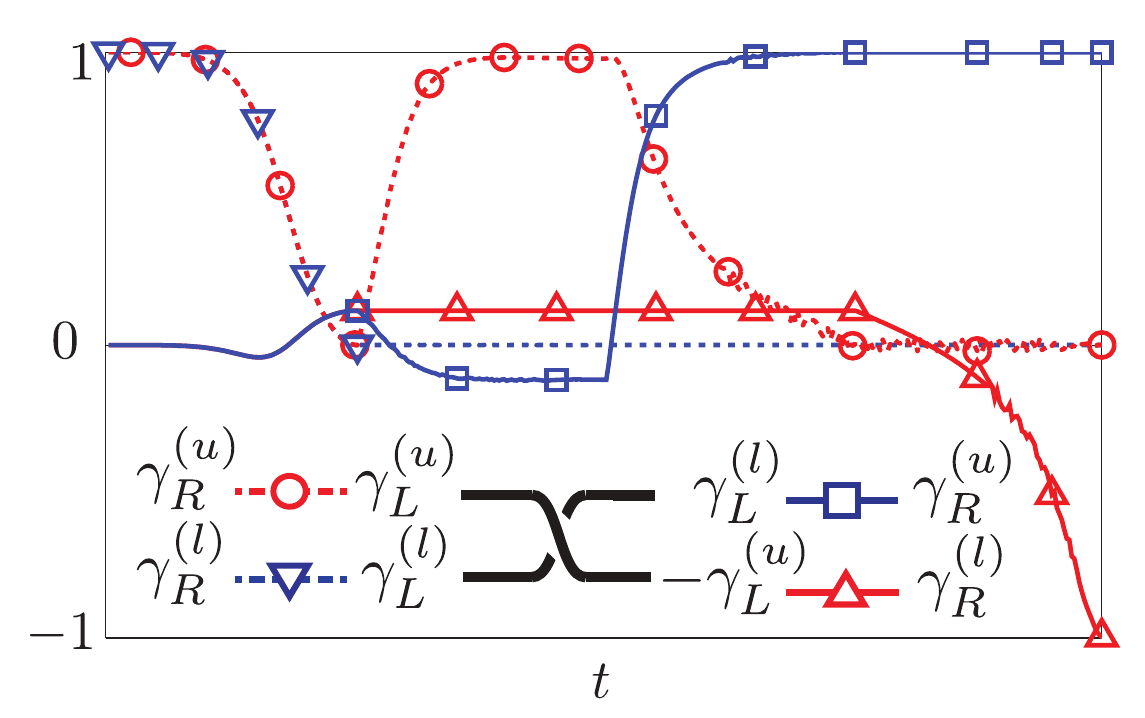}
\end{center}
\caption{Adiabatic braiding protocol in the presence of a harmonic trap with $V_{t}=J$: Adiabatic evolution
of the Majorana correlation functions $\langle i\gamma_{L}^{(u)}\gamma
_{R}^{(u)}\rangle$ (red, $\circ$), $\langle i\gamma_{L}^{(l)}\gamma_{R}%
^{(l)}\rangle$ (blue, $\bigtriangledown$), $\langle i\gamma_{L}^{(l)}%
\gamma_{R}^{(u)}\rangle$ (blue, $\square$), and $\langle i\gamma_{L}%
^{(u)}\gamma_{R}^{(l)}\rangle$ (red, $\bigtriangleup$) during the braiding
protocol with errors $\delta_K$ in the local operations (see main text) for two
non-ideal quantum wires of length $L=40$ with $|\Delta|=1.2J, \mu = 0$ and $\delta_K=0.05$.
}%
\label{fig:harmonic_braiding}
\end{figure}

In Fig.~\ref{fig:harmonic_density_modes} we compare the numerical results with and without the harmonic
potential. We consider a potential with strength $V_{t}=J$ for quantum wires of $N=40$ sites,
$|\Delta|=1.2J$, $\mu=0$, and $\delta_K = \delta_\perp = \delta_v =0.05$. This potential is strong enough to have a visible effect on the density distribution 
$n(j)=\langle a_{j}^{\dagger}a_{j}\rangle$, and also changes the MF wave function $|v_{j}|$ defined via $\gamma_{L}=\sum_{j}%
v_{j}c_{2j-1}$. In Fig.~\ref{fig:harmonic_braiding} we show the evolution of the correlations between
MFs during the braiding protocol, concluding that the harmonic potential does not affect the results of
the braiding. Note also that in the case of a harmonic trap, the correlation functions $\langle i \gamma_L^{(l)}\gamma_R^{(u)}\rangle$ and $\langle i \gamma_L^{(u)}\gamma_R^{(l)}\rangle$ have larger $(\sim 0.1)$ values in the middle of the protocol as compared to the case with no harmonic confinement (see Fig.~\ref{fig:nonperfect} of the main text) where this difference cannot be distinguished. The non-zero values of these correlations are due to the overlap of the wave functions (coefficients $v_j$) of the evolving Majorana zero modes with those at the beginning of the protocol. As it is illustrated in Fig.~\ref{fig:harmonic_density_modes}, a shallow harmonic trap increases the extension of the Majorana zero modes and therefore leads to larger overlaps.  

From the discussion of the last two subsections we can conclude that the braiding protocol is insensitive to the class of errors most likely to dominate in an experiment as long as the two Majorana wave functions are spatially well separated, and the protocol is performed on a time scale which satisfies adiabaticity. This resilience against error can be intuitivly understood by recalling that the protocol is based on extracting and re-inserting one physical fermion. 

\subsection{Demonstration of non-Abelian statistics of MFs via parity measurement}

\begin{figure}[tbp]
\includegraphics[width=0.95\columnwidth]{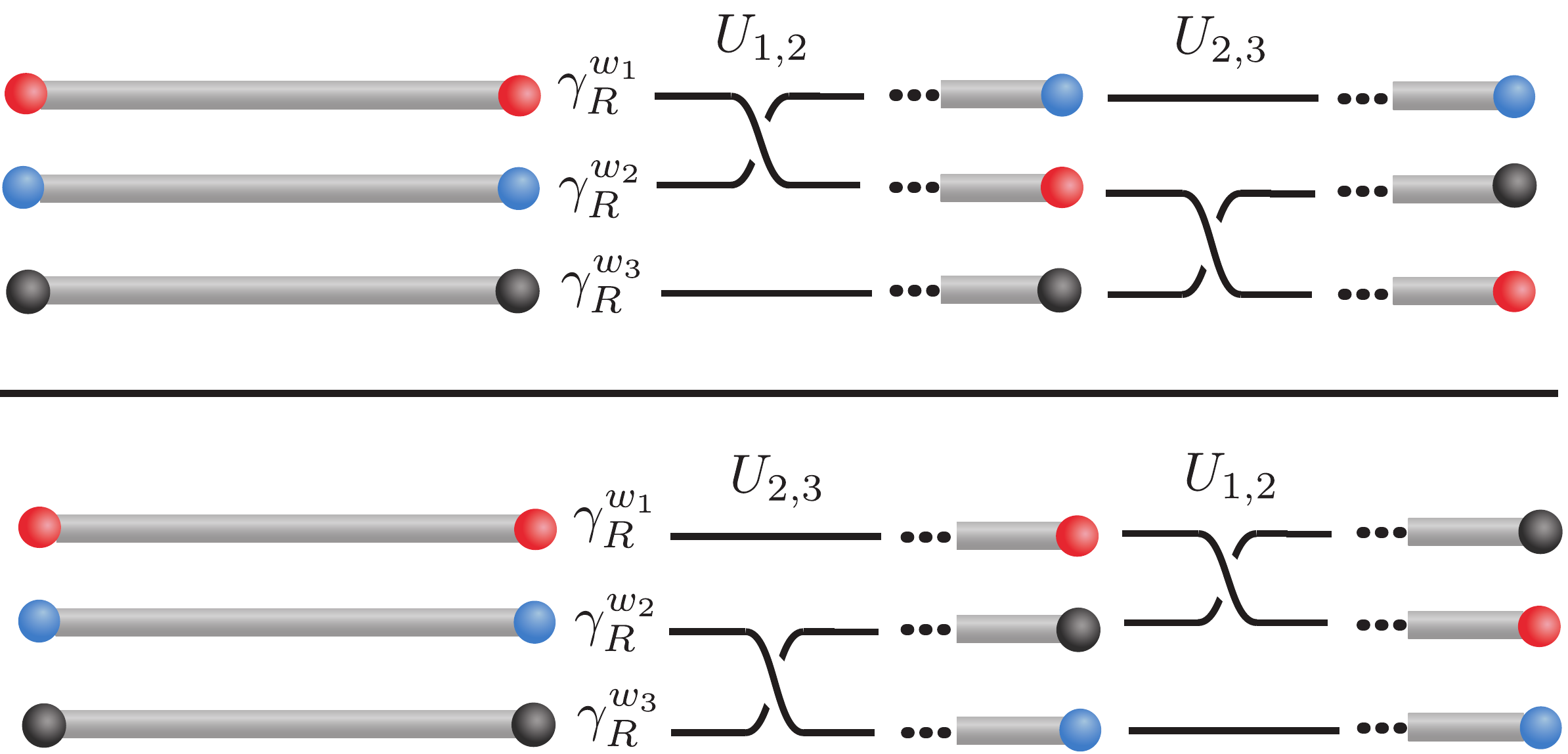}
\caption{A setup of three Kitaev wires, $w_1,w_2,w_3$ with Majorana modes $\gamma_L^{w_j}, \gamma_R^{w_j}$ located at the left and right ends of each wire, respectively. The braiding of neighbouring Majorana modes $U_{1,2} = {\rm exp}[\pi \gamma_R^{(w1)} \gamma_R^{(w2)}/4 ] $ and $U_{2,3} = {\rm exp}[\pi \gamma_R^{(w2)} \gamma_R^{(w3)} /4]$ do not commute, i.e. $U_{1,2}U_{2,3} \ne U_{2,3}U_{1,2}$. This non-commutativity can be probed with a parity measurement of a single wire (see text). }
\label{fig:show_braiding}
\end{figure}

 The non-abelian nature of MFs manifests itself as the non-commutativity of the braiding operations between them. Here we propose a simple setup to demonstrate this non-commutativity via a single parity measurement of an individual wire. For this purpose, we consider a system of three Kitaev wires (labelled $w1,w2,w3$), each supporting Majorana modes ($\gamma_L^{(w_j)}, \gamma_R^{(w_j)}$) located at the left and right ends of each wire, as shown in Fig.~\ref{fig:show_braiding}.  The two braiding operations $U_{1,2} = {\rm exp}[\pi \gamma_R^{(w1)} \gamma_R^{(w2)}/4 ] $ and $U_{2,3} = {\rm exp}[\pi \gamma_R^{(w2)} \gamma_R^{(w3)} /4]$ act to interchange the Majorana fermions $\gamma_R^{(w1)} \gamma_R^{(w2)}$ and $\gamma_R^{(w2)} \gamma_R^{(w3)}$, respectively, at the ends of neighbouring wires. The non-abelian statistics of the Majorana fermions implies that these braiding operations do not commute: If we begin in the even parity ground state of each wire, $|+_{w1}+_{w2}+_{w3}\rangle$, then braiding $U_{1,2} U_{2,3} $ and $U_{2,3} U_{1,2} $ results in
\begin{align}
&U_{1,2}U_{2,3} |+_1+_2+_3\rangle = \notag \\ &\frac{1}{2} \left(|+_1+_2+_3\rangle-|+_1-_2-_3\rangle-|-_1-_2+_3\rangle+|-_1+_2-_3\rangle \right)  \notag \\
&U_{2,3}U_{1,2} |+_1+_2+_3\rangle = \notag \\ & \frac{1}{2} \left(|+_1+_2+_3\rangle-|+_1-_2-_3\rangle-|-_1-_2+_3\rangle-|-_1+_2-_3\rangle \right),
\end{align} 
which differ in the sign of the last term.  While this difference in sign could be impractical to measure, a simpler demonstration of the non-Abelian statistics can be done by checking the non-commutativity of pairs of consecutive braids $U_{1,2}U_{2,3}$ and $U_{2,3}U_{1,2}$:
\begin{align}
&U_{1,2}U_{2,3}^2U_{1,2} |+_1+_2+_3\rangle =  |+_1-_2-_3\rangle \notag\\
&U_{2,3}U_{1,2}^2U_{2,3}|+_1+_2+_3\rangle = |-_1-_2+_3\rangle.
\end{align} 
The difference can now be easily distinguished by a single parity measurement of the first or third wire, directly showing the effect of the non-abelian statistics of MFs.

%%%%%%%%%%%%%%%%%%%%%%

 %%%%%%%%%%%%%%%%%%%%%%%%%%%%%%%%%%%%%%%%%%%%%%
\section{Topological qubits and gates in a network of quantum wires} 
In the previous subsections we presented an extended discussion of the realisation and braiding of MFs in an optical lattice setup. Now we proceed by defining a qubit subspace from a network of Kitaev wires and describing how braiding the resulting MFs implements quantum gates. 

\begin{figure}[t!]
	\begin{center}
		\includegraphics[width=0.45\textwidth]{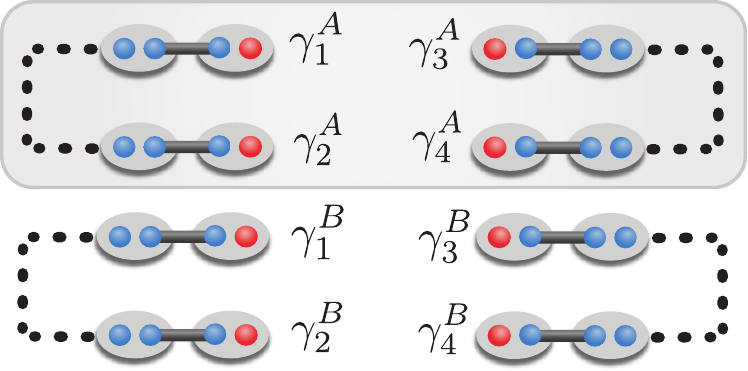}
		\caption{Definition of qubits in an optical lattice setup. Each wire is formed into a U-shape, such that the Majorana fermions  (i.e. $\hat{\gamma}_1^A$ and $\hat{\gamma}		_2^A$) are located next to each other on the lattice. This setup is scalable and allows for Majorana fermions to be braided. A single qubit, labelled as $A(B)$, is defined 	on two wires with four Majorana fermions $\hat{\gamma}_i, i=1-4$.}
		\label{fig:define_qubits_braiding}
	\end{center}
\end{figure}

Due to their non-local structure and their topological origin, the MFs are intrinsically robust against local perturbations. This renders the ground state subspace of the Kitaev wire an ideal system for storing quantum information. Note, that one Kitaev wire is not enough to store a qubit. The two ground states, $|+\rangle$ and $|-\rangle$ have different fermionic parity, and their superposition is forbidden due to superselection rules. This problem can be overcome by using two wires with four Majorana fermions to define a qubit basis, as shown in Fig.~\ref{fig:define_qubits_braiding}. In order to perform quantum gates via braiding efficiently, we realize each of the two wires in a U-shape, and denote the two MFs on the left wire as $\gamma_{1,2}^A$ while the two MFs on the right wire are labeled $\gamma^A_{3,4}$. We denote by $|-\rangle_{x}^A$ the odd-parity ground state of the left ($x =L $) and right ($x=R$) chain, respectively. Then, we define the local qubit basis for qubit $A$, $|\bar{0}\ra$ and $|\bar{1}\ra$ via the two odd-parity states

\begin{align}
	\label{eq:define_logic}
		|\bar{0}\ra \equiv |+\rangle^A_L\otimes |-\rangle^A_R =  f_L^{A\dagger}  |-\rangle_L^A\otimes |-\rangle_R^A  \nonumber \\
		|\bar{1}\ra\equiv |-\rangle^A_L\otimes |+\rangle^A_R =  |-\rangle_L^A  \otimes f_R^{A\dagger}|-\rangle_R^A ,
\end{align}
where $f_L^{A\dagger} = \gamma_1^A-i \hat{\gamma}_2^A$ and $f_R^{A\dagger} = \gamma_3^A-i \gamma_4^A$. This setup can be readily scaled up to $N$ qubits, as shown in Fig.~\ref{fig:define_qubits_braiding} for the case of $N=2$, where the two qubits are labelled $A$ and $B$. For an alternative way to define an N-qubit space with a definite parity see \cite{Georgiev06}.

 From the setup depicted in Fig.~\ref{fig:define_qubits_braiding} we conclude that the protocol allows us to braid MFs on any two adjacent ends of one or two wires. This implies we can realize the unitaries $\hat{U}_{12}^{\alpha \alpha}$, $\hat{U}_{13}^{\alpha \alpha}$, $\hat{U}_{21}^{\alpha \beta}$ and $\hat{U}_{43}^{\alpha \beta}$, where $\alpha$ and $\beta$ label two adjacent wires and $\hat{U}_{ij}^{\alpha \beta}=(1-\hat{\gamma}_i^\alpha\hat{\gamma}_j^\beta)/\sqrt{2}$. Note that braiding Unitaries $  \hat{U}_{34}^{\alpha \alpha} ,  \hat{U}_{24}^{\alpha \alpha}$ are, up to a phase, equivalent to an overall phase $ \hat{U}_{12}^{\alpha \alpha} ,\hat{U}_{13}^{\alpha \alpha}$. These unitaries will result in single qubit operations as well as two-qubit operations on neighboring qubits. First, let us consider braids on one wire only; these will result in single qubit operations. As we show in Appendix \ref{sec:append_braiding}, the unitaries $\hat{U}_{12}^{\alpha \alpha}$ and $\hat{U}_{13}^{\alpha \alpha}$ realize the single qubit Pauli gates ($\hat{X},\hat{Y},\hat{Z}$) and the Hadamard ($\hat{H}$) gate, via

\begin{eqnarray}
		\hat{Z}_\alpha \otimes \mathbb{1} &=& \hat{U}_{12}^{\alpha \alpha} \hat{U}_{12}^{\alpha \alpha} \nonumber \\
		%\mathbb{1}  \otimes Z&=& U_{5,6} U_{5,6} \nonumber \\
		\hat{X}_\alpha \otimes \mathbb{1} &=& \hat{U}_{13}^{\alpha \alpha} \hat{U}_{13}^{\alpha \alpha} \hat{U}_{12}^{\alpha \alpha} \hat{U}_{12} \nonumber \\
		%\mathbb{1}  \otimes X&=& U_{5,7} U_{5,7} U_{7,8} U_{7,8} \nonumber \\
		\hat{H}_\alpha \otimes \mathbb{1} &=& \hat{U}_{13}^{\alpha \alpha}   \hat{U}_{12}^{\alpha \alpha} \hat{U}_{12}^{\alpha \alpha}, 
		%\mathbb{1}  \otimes HZ&=& U_{5,7}
	\end{eqnarray}
and $\hat{Y}_\alpha =\hat{Z}_\alpha\hat{X}_\alpha$. To complete all single qubit rotations it is also necessary to implement a $\pi/8$ phase gate. However, as shown in \cite{Ahlbrecht09} it is not possible to realise this gate due to the form of the braiding unitary (Eq.~\eqref{eq:braiding_unitary}).   

 Secondly, we consider braids on neighbouring wires $\hat{U}_{21}^{\alpha \beta}$. Alone this unitary takes us out of the logical qubit subspace. However, a combination of these unitaries with the braids involving one wire only result in the SWAP gate: 
\begin{align}\label{eq:swap}
\hat{U}_{\mathrm{SWAP}}= \left[\hat{U}_{21}^{AB} \hat{U}_{43}^{AB}\right] \left[ \hat{U}_{12}^{AA} \hat{U}_{34}^{AA}\hat{U}_{12}^{BB} \hat{U}_{34}^{BB} \right]\left[ \hat{U}_{21}^{AB} \hat{U}_{43}^{AB}\right].
\end{align}
Unitaries that are written in one square bracket can be carried out simultaneously, so that the implementation of the SWAP can be achieved in three steps only. Eq. \eqref{eq:swap} can be easily verified by multiplying the $16 \times 16$ matrices that describe the effect of the braiding on the physical space $\{|x\rangle^A_L|x\rangle^A_R |x\rangle^B_L|x\rangle_R^ B\}_{x = +,-}$.

While these braiding operations form the basis of topologically protected gates, braiding operations of Ising anyons alone are not sufficient for UQC \cite{Freedman03, Nayak96}. With the setup described above, a $\pi/8$-phase gate and an entangling gate complete the gate set for UQC.
%========================================================================================
%========================================================================================
\section{Mapping between Topologically Protected and Unprotected Space}
\label{sec:interface}

In this section we introduce an efficient, robust and reversible mapping that allows for an interface between the topologically protected qubit space to a topologically unprotected space. This mapping provides a platform for: (i) initialisation of a wire in a desired parity state, (ii) measurement and (iii) the implementation of the missing gates required for universal quantum computation. 
%========================================================================================
 As was the case for the braiding protocol, the mapping protocol is based on the toolbox available for optical lattices introduced in Section~\ref{sec:coldatom_implementation}, in particular, the ability to address individual sites and links. The objective is to map the non-local fermions of each qubit ($\gamma^A_1+i\gamma^A_2$ and $\gamma^A_3+i\gamma^A_4$ in Fig.~\ref{fig:define_qubits_braiding}) to a local, physical fermion on an additional site on the lattice. This locality is the key for state initiation, measurement and for the construction of the missing quantum gates. 

	\begin{figure}[t!]
	\begin{center}
		\includegraphics[width=0.45\textwidth]{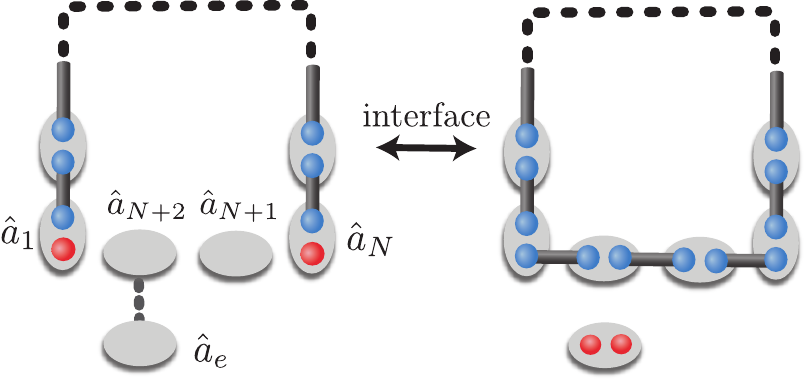}
		\caption{Schematic of the mapping protocol. {\bf Left:} At $t=0$ the chain is open and includes $N$ sites, coupled via hopping and pairing (shown as a grey link). The external sites, corresponding to the operators $\ha_{N+1}, \ha_{N+2}, \ha_e$ are completely decoupled from the chain, but the site of operator $\ha_{N+2}$ is coupled to that of $\ha_e$ via hopping only (shown as a grey dashed link). {\bf Right:} At $t=t_f$ the chain is closed and includes $N+2$ sites. There is one external site which will remain occupied or empty depending on the initial parity of the open chain. Closing the chain across two sites introduces a crucial geometric asymmetry, the purpose of which is discussed in the main text. }
		\label{fig:schematic_close}
	\end{center}
	\end{figure}
	
	\subsection{Basic idea}
\label{sec:basic_idea}
In Fig.~\ref{fig:schematic_close} we show a minimal setup for carrying out this protocol. The Kitaev Hamiltonian $H_K$ (given by Eq.~\eqref{eq:Kitaev} ) is realized on $N$ sites that are arranged in a U-shape configuration. We denote by $|+/-\rangle$ the even and odd parity ground states of $H_K$. The two ends of the wire, $1$ and $N$ must be separated by at least two empty sites which we denote by $N+1$ and $N+2$. These sites are initially decoupled from the chain. Further, the site $N+2$ is coupled via hopping to a site denoted by $e$, which we call the external site, and which is initially empty. The site $e$ can host one fermion, with creation operator $\hat a^\dagger_e$. The geometric asymmetry in this setup ensures an energy gap between the two lowest lying states and the higher excited states throughout the protocol. This gap will set the time scale of the adiabatic protocol, as will be described in detail in the following subsection. An alternative setup would be completely geometrically symmetric, however asymmetric in coupling parameters.

Consider for the moment only the open chain and the sites $N+1$ and $N+2$, and assume that we couple these two sites adiabatically with the open chain and with each other, such that in the end we obtain a realization of the Kitaev Hamiltonian on a closed ring of length $N+2$, realising the `closed chain' shown in Fig.~\ref{fig:define_Kitaev}b.  If we allow only parity preserving operations during the closing of the chain, the adiabatic theorem implies that the odd parity ground state $|-\rangle$ is mapped to the (unique) ground state $|g\rangle$ of the closed chain, while the even parity ground state $|+\rangle$ is mapped to an excited state of the system $\tilde{a}_\nu^\dagger|g\ra$ (see the discussion in Sec.~\ref{sec:theoretical_model}). 

Now, the external site, initially coupled to site $N+2$, comes into play. During the adiabatic passage from the open to the closed chain, the hopping between the sites $e$  and $N+2$ is switched off adiabatically; at the end of the evolution the site $e$ is decoupled from the closed chain. As we explain in detail below, this process can be engineered such that 	 \begin{eqnarray}
	 \label{eq:state_evolution_desired}
		 |-\ra&\rightarrow &|g\ra\otimes|\Omega_e\rangle \nonumber \\
		 |+\ra&\rightarrow &|g\ra\otimes\hat{a}_e^\dagger |\Omega_e\rangle,
	 \end{eqnarray}
where an empty site $j$ is denoted by $|\Omega_j\rangle$. Eq.~\eqref{eq:state_evolution_desired} implies the odd (even) parity ground state of the open Kitaev chain is mapped to an empty (occupied) external site in a reversible way.

\subsection{Mapping Protocol Hamiltonian}

 \begin{figure*}[t!]
	\begin{center}
	%\hspace{-4mm}
		\includegraphics[width=0.8\textwidth]{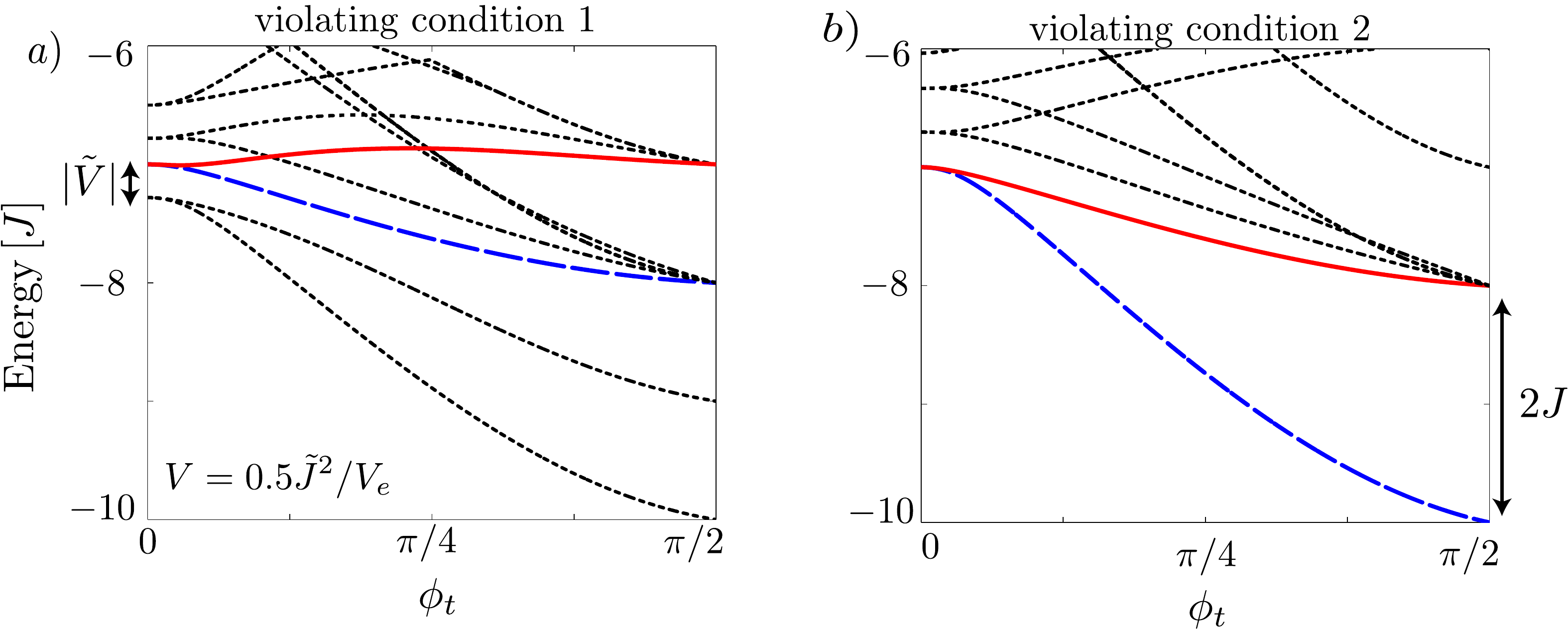}
		\caption{Energies of the lowest energy eigenstates of the Hamiltonian in Eq.~(\ref{eq:total_ham_interface}) as a function of the adiabatic parameter $\phi_t \in [0,\pi/2]$. At $\phi_t=0$ there are two degenerate ground states with odd and even parity respectively. Parameters: $\Delta=J, N=8, \tilde{J}=J$. {\bf Panel~$\mathbf{a)}$} $V_e=J, V=0.5 \tilde{J}^2/V_e=J/2$. The initial states of the system are no longer the ground states of the system rather they are excited states with energy $|\tilde{V}|=(3-\sqrt{17})J/4$  as compared to the ground state (see Eq.~\eqref{eq:energy_symmstate}). The evolution will be $|-\ra\otimes|\Omega_{N+1,N+2,e}\rangle \rightarrow \tilde{a}_{\nu}^\dagger|g\ra\otimes\hat{a}_e^\dagger |\Omega_e\rangle$ (shown in blue-dashed)  and $|+\ra\otimes|\Omega_{N+1,N+2,e}\rangle \rightarrow \tilde{a}_{\nu}^\dagger |g\ra\otimes|\Omega_e\rangle$ (shown in red). Additional eigenstates are shown as black-dotted lines. The excited states of the chain are degenerate, as seen by the intersection of states at $\phi_t=\pi/2$. {\bf Panel~$\mathbf{b)}$} $V_e=3J, V= 2\tilde{J}^2/V_e = 2J/3$. It is no longer energetically favourable for a particle to occupy the site $e$, therefore the evolution will be $|-\ra\otimes|\Omega_{N+1,N+2,e}\rangle \rightarrow |g\ra \otimes |\Omega_e\ra $ (shown in blue-dashed) and $ |+\ra\otimes|\Omega_{N+1,N+2,e}\rangle \rightarrow \tilde{a}_{\nu}^\dagger |g\ra\otimes|\Omega_e\rangle$ (shown in red). Again, the degeneracy of the excited states of the chain gives an intersection of states at $\phi_t=\pi/2$.}
		\label{fig:energy_diagram}
	\end{center}
	\end{figure*}
\begin{figure}[t!]
	\begin{center}
		\includegraphics[width=0.4\textwidth]{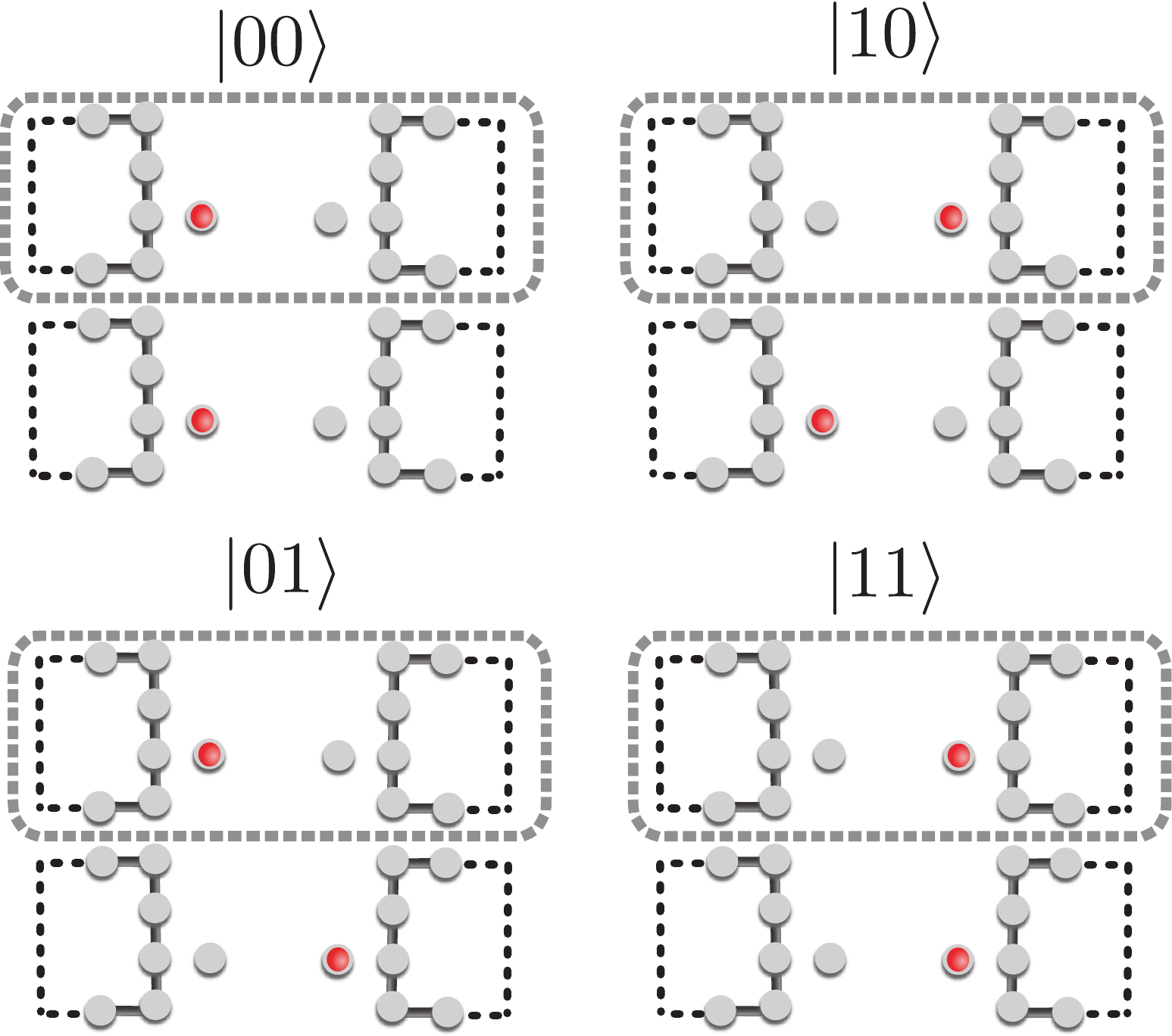}
		\caption{The result of the mapping  on two qubits, each defined according to Eq.~\eqref{eq:define_logic}. A single qubit is shown enclosed in a grey dashed line. At the end of the mapping  and external site is occupied (shown as a red site) on the left (if the qubit was in the state $|\bar{0}\rangle$) or on the right (if the qubit was in the state $|\bar{1}\rangle$).}
		\label{fig:define_qubit_image}
	\end{center}
	\end{figure}

Let us now describe this mapping protocol in detail. The protocol can be carried out using only operations on the sites $N+1$, $N+2$ and  $e$ and the associated links. It requires the ability to switch on/off (i) the hopping between the site $N+2$ and the external site, $H^{(h,\tilde{J})}_{N+2,e} = \tilde{J} \hadag_{N+2}\hat{a}_{e}+{\rm h.c.}$, (ii) the couplings $H^{(K,J,\Delta)}_{k,l} = -Ja^{\dagger}_{k}a_{l} + \Delta a_{k}a_{l} +{\rm h.c.}$ between any two adjacent sites $k,l \in \{1, N, N+1, N+2\}$ and (iii) the local potentials $H^{(V)}_k = V \ad_k a_k$ on the sites $k=N+1, N+2$.  Again, we model the adiabatic passage via two continuous and monotonic time-dependent functions $C_t,S_t: [0, t_{f}] \rightarrow [0,1]$, with $C(0) = 1, C(t_{f}) = 0$ and $S(0) = 0, S(t_{f})=1$. 	The Hamiltonian of the mapping is then given by

	\begin{align}
	\label{eq:total_ham_interface}
		\hat{H}_{I} &= H_K +H^{(V_e)}_{e} +  C_t \big[H^{(h,\tilde{J})}_{N+2,e}+H^{(V)}_{N+2}+H^{(V)}_{N+1} \big]\nonumber\\
		& + S_t \big[ \hat{H}^{(K,J,\Delta)}_{N,N+1}+\hat{H}^{(K,J,\Delta)}_{N+1,N+2}+\hat{H}^{(K,J,\Delta)}_{N+2,1}\big], 
			\end{align}
			where $ H_K$ is the Hamiltonian of the open Kitaev chain with $N$ sites (Eq.~\eqref{eq:Kitaev}). At $t=0$ the ground state of the Hamiltonian $H_I$ will be a tensor product of the ground states of three decoupled systems: the Kitaev wire, the decoupled site $N+1$ and the sites $N+2$ and $e$ that are coupled together by hopping $\tilde{J}$. At this time, the even(odd) parity ground states are $|+(-)\rangle \otimes |\Omega_{N+1}\rangle \otimes|\Omega _{N+2,e}\rangle$, provided that:
			\begin{eqnarray}
			\label{eq:energy_symmstate}
			V&>&0, \nonumber \\
			\tilde{V}=\big(V+V_e-\sqrt{(V-V_e)^2+4 \tilde{J}^2}\big)/2 &>&0.
			\end{eqnarray}
			Here, $V$ is the energy of a particle occupying the site $N=1$, and $\tilde{V}$ is the energy of a single particle occupying the eigenstate $(a^\dagger_{N+2}+a^\dagger_{e})|\Omega_{N+2,e}\rangle/\sqrt{2}$, thus the above conditions ensure that the ground state on these sites is the vacuum. 
The above conditions are satisfied with a potential $V$ satisfying:
\begin{equation}
\label{eq:energy_condition1}
V>\frac{ \tilde{J}^2}{V_e}>0, \hspace{0.3cm} \mathrm{(condition \;1)}
\end{equation}
Violating condition 1 would alter the form of the ground state; the consequence of this violation will be discussed further below. Throughout the evolution from $0$ to $t_f$ the adiabatic theorem ensures that the system stays in the corresponding eigenstate, as long as the energies within a given parity subspace are non-degenerate. 
If we further impose a second condition 
	\begin{eqnarray}
		0&<&V_e< 2J,  \hspace{0.3cm} \mathrm{(condition \;2)}
		\label{eq:energy_condition2}
	\end{eqnarray}
we ensure that the odd parity ground state $|-\rangle$ transfers to the state $|g\rangle$ and an empty external site, while the even parity ground state $|+\rangle$ transforms to $|g\rangle$ and an occupied external site.  Thus, by tuning the Hamiltonian adiabatically we obtain $\la \hat{a}_{e}^\dagger\hat{a}_{e}\ra = 1 (0)$ if the original parity of the chain was even (odd), as shown in Fig.~\ref{fig:define_qubit_image}. 
\begin{figure*}[t!]
	\begin{center}
	%\hspace{-4mm}
		\includegraphics[width=0.8\textwidth]{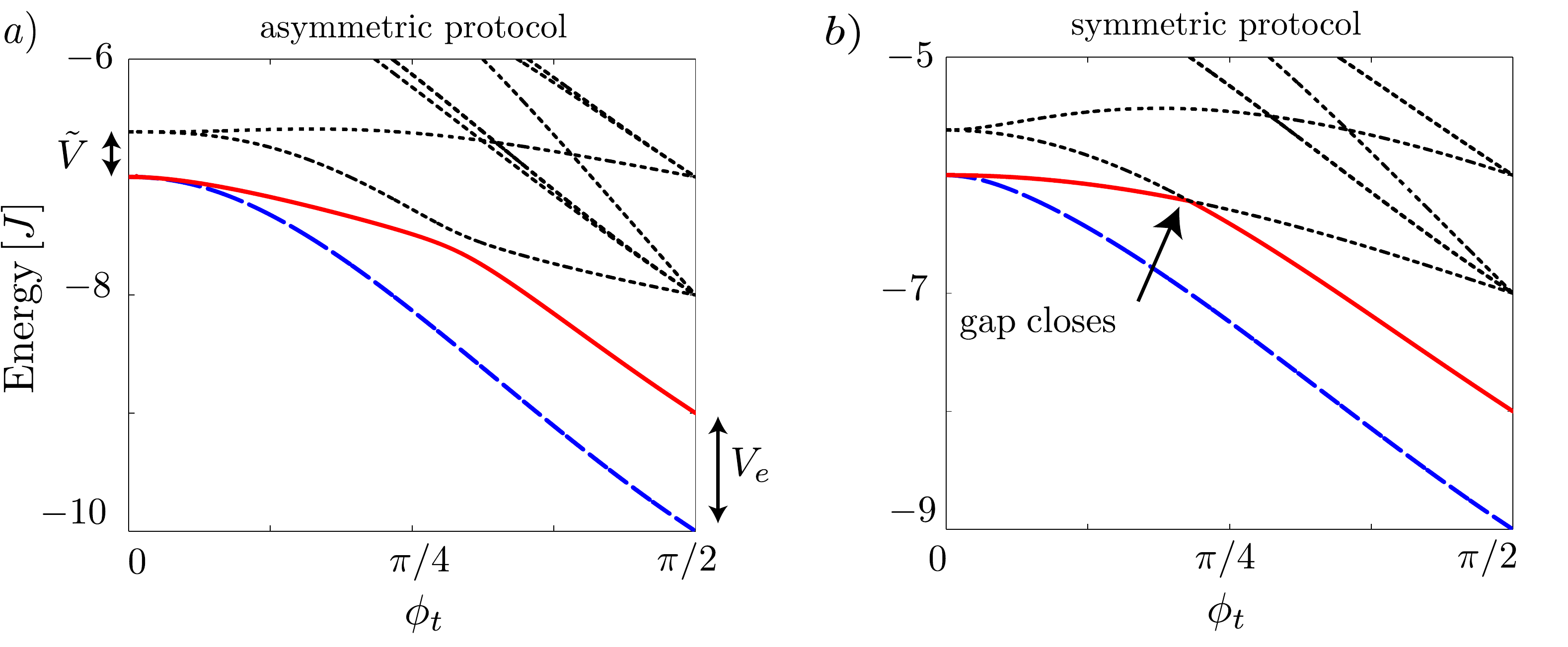}
		\caption{Energies of the lowest energy eigenstates of the Hamiltonian in Eq.~(\ref{eq:total_ham_interface}) as a function of the adiabatic parameter $\phi_t \in [0,\pi/2]$, with parameters: $\Delta=J, N=8, \tilde{J}=J, V_e=J,V=2 \tilde{J}^2/V_e=2J$.  At $\phi_t=0$ there are two degenerate ground states with odd and even parity respectively, in this parameter regime the next excited states are gapped by $\tilde{V}=(3-\sqrt{5})J/2$ (see Eq.~\eqref{eq:energy_symmstate}) . {\bf Blue, Dashed:} Evolution of the state $|-\ra\otimes|\Omega_{N+1,N+2,e}\rangle \rightarrow |g\ra\otimes|\Omega_e\rangle$.  {\bf Red:} Evolution of the state $|+\ra \otimes|\Omega_{N+1,N+2,e}\rangle \rightarrow |g\ra\otimes\hat{a}_e^\dagger |\Omega_e\rangle$.  At $\phi_t = \pi/2$ this state has an energy offset $V_e$ compared to the ground state. {\bf Black, Dotted:}. Higher energy states of the system. \newline Panel $a)$ Asymmetric closing, as shown in Fig.~\ref{fig:schematic_close}.  Panel $b)$ Closing protocol with a symmetric setup, obtained by omitting site $N+1$ in Fig.~\ref{fig:schematic_close}. In this case of no asymmetry, the gap between the lowest lying states, and the higher excited states closes.}
		\label{fig:energy_diagram}
	\end{center}
	\end{figure*}

	%%%%%%%%%%%%%%%	
	
Despite its simple form, the eigen-energies of $H_I(t)$ cannot be represented in a compact analytic form, even in the special case of $J=\Delta$. Thus, to show the evolution of each state under the Hamiltonian $H_I(t)$ subject to condition 1 and 2, we have carried out a detailed numerical analysis. A summary of the results is presented in Fig.~(\ref{fig:energy_diagram}), where we present the evolution of the lowest lying eigen-energies  of the Hamiltonian Eq.~(\ref{eq:total_ham_interface}). We parametrize the time evolution via $C_t = \cos{\phi_t}$, $S_t =\sin{\phi_t}$ with the adiabatic parameter $\phi_t$ that changes smoothly from $0$ to $\pi/2$, as $t$ changes from $0$ to $t_f$.  In Fig.~\ref{fig:energy_diagram} a we present the evolution for the setup described above for the case of an open chain of $N=8$ sites, $J = \Delta = V_e = \tilde J$ and  $V=2\tilde{J}^2/V_e = 2J$. As $\phi_t$ is changed adiabatically from $0$ to $\pi/2$ the energies of two states, $|+\rangle$ and $|-\rangle$ begin to split resulting in the evolution 
\begin{eqnarray}
\label{eq:desired_evolution}
|-\ra\otimes|\Omega_{N+1,N+2,e}\rangle &\xleftrightarrow{\text{mapping }}&|g\ra\otimes|\Omega_e\rangle\nonumber \\
 |+\ra \otimes|\Omega_{N+1,N+2,e}\rangle&\xleftrightarrow{\text{mapping }}& |g\ra\otimes\hat{a}_e^\dagger |\Omega_e\rangle.
 \end{eqnarray} The minimal gap between the energy of the second (red line) and third (black line) lowest lying states, $\Delta E_{2,3}$ gives a measure for the speed of the adiabatic process. From Fig.~\ref{fig:energy_diagram} b it becomes clear why we need to close the chain via two external sites: The asymmetry introduced in our setup is fundamental for the existence of a finite gap between the second (red line) and third (black line) lowest lying states. 

 Now we consider the effect of violating the conditions $1$ and $2$ in Eq.~(\ref{eq:energy_condition1}) and Eq.~(\ref{eq:energy_condition2}).  If we violate condition 1, we choose $0<V<\frac{ \tilde{J}^2}{V_e}$. The even and odd parity ground states of $H_I$ at $\phi_t=0$ will have now have one particle present in the symmetric eigenstate of the sites $N+1$ and $e$. By contrast, our initial states, for which these sites are empty, are now excited states with an energy $|\tilde{V}| = |(V+V_e-\sqrt{(V-V_e)^2+4 \tilde{J}^2})/2| >0$ compared to the ground state. These states will evolve as excited states and we obtain 
 \begin{eqnarray}
 \label{eq:violate_condition1}
|-\ra\otimes|\Omega_{N+1,N+2,e}\rangle &\xleftrightarrow{\text{mapping }}&\tilde{a}_{\nu}^\dagger|g\ra\otimes\hat{a}_e^\dagger |\Omega_e\rangle \nonumber \\
|+\ra\otimes|\Omega_{N+1,N+2,e}\rangle &\xleftrightarrow{\text{mapping }}&\tilde{a}_{\nu}^\dagger |g\ra\otimes|\Omega_e\rangle
 \end{eqnarray}
  At the end of the mapping  these states are largely degenerate, with $\nu \in [1,N+2]$ (see the discussion in Sec.~\ref{sec:theoretical_model}). This will cause a mixing of states at the end of the evolution and the process will not be reversible. We illustrate this for the case of $V=0.5 \tilde{J}^2/V_e$ in Fig.~\ref{fig:energy_diagram} a, where at $\phi_t=\pi/2$ one can explicitly see the crossing of our initial states (in blue and red) with the other degenerate excited states of the chain. 
  
  On the other hand, if we violate condition 2 by taking the potential $V_e$ on the external site to be too large, then it is no longer favourable for the external site to become occupied. While the odd parity state evolves to the ground state as desired, the even parity site evolves to an excited state of the chain, with the external site remaining empty,  
\begin{eqnarray}
 \label{eq:violate_condition2}
|-\ra\otimes|\Omega_{N+1,N+2,e}\rangle &\xleftrightarrow{\text{mapping }}& |g\ra \otimes |\Omega_e\ra \nonumber \\
|+\ra\otimes|\Omega_{N+1,N+2,e}\rangle &\xleftrightarrow{\text{mapping }}&\tilde{a}_{\nu}^\dagger |g\ra\otimes|\Omega_e\rangle.
 \end{eqnarray}
 Again, due to the large degeneracy of the state $\tilde{a}_{\nu}^\dagger |g\ra$ this evolution will not be reversible.
As an example, we show in Fig.~\ref{fig:energy_diagram} b the evolution for $V_e = 3J$.

As we have shown above, the mapping  protocol allows us to map the parity of the Kitaev chain to the occupation of a physical site, and the adiabaticity ensures that the protocol is reversible. As we will discuss below, this mapping  between a topologically protected and unprotected space has several applications for quantum computation. As a first application, note that the mapping  can be used to initialize and measure an arbitrary qubit state: To each of the two wires forming one logical qubit via $|\bar 0 \rangle = |+\rangle_L|-\rangle_R$, $|\bar 1 \rangle = |-\rangle_L|+\rangle_R$ we associate an external site, $e_L$ and $e_R$ respectively, which can each host one fermion, $\ad_{e_{L}},\ad_{e_{R}}$. Then, applying the mapping  protocol on both wires simultaneously, we see that 
\begin{align}
\label{eq:interface_qubits}
|\bar 0 \rangle &= |+\rangle_L|-\rangle_R \xleftrightarrow{\text{mapping }}  \ad_{e,L}|\Omega_e\rangle_L \otimes|\Omega_{e,R}\rangle\\
|\bar 1 \rangle &=  |-\rangle_L|+\rangle_R  \xleftrightarrow{\text{mapping }}  |\Omega_e\rangle_L \otimes \ad_{e,R}|\Omega_{e,R}\rangle.
\end{align}
This setup scales naturally to $N$ qubits; the logical subspace for $N=2$ is shown schematically in Fig.~\ref{fig:define_qubit_image}.

In the following section we discuss the experimental errors which arise in implementing the Hamiltonian in Eq.~(\ref{eq:total_ham_interface}). 
%========================================================================================
%========================================================================================
\subsection{Effect of imperfections in a cold atom setup}

%========================================================================================
 
When considering the physical implementation of the Hamiltonian Eq.~(\ref{eq:total_ham_interface}), we consider the following set of experimentally relevant errors: (i) a non-ideal Kitaev chain ($\Delta \ne J, \mu \ne 0$ in Eq.~(\ref{eq:Kitaev_partII})), with deviations on the order of $\Delta -J\sim 0.1 J, \mu \sim 0.1 J$ (ii) local fluctuations in the lattice ($\mu_i = \mu \pm \mu_r$, where $\mu_r \in [-0.1J,0.1J] $ is a random fluctuation on each site), (iii) different lasers may be tuned at different timings (implying that the functions $C_t, S_t$ can vary independently in time while still satisfying $C(0) = 1, C(t_{f}) = 0$ and $S(0) = 0, S(t_{f})=1$), (iv) the lasers focused on an individual site are not perfect, giving $10\%$ laser intensity on neighbouring sites.  

\begin{figure}[t!]
	\begin{center}
		\includegraphics[width=0.4\textwidth]{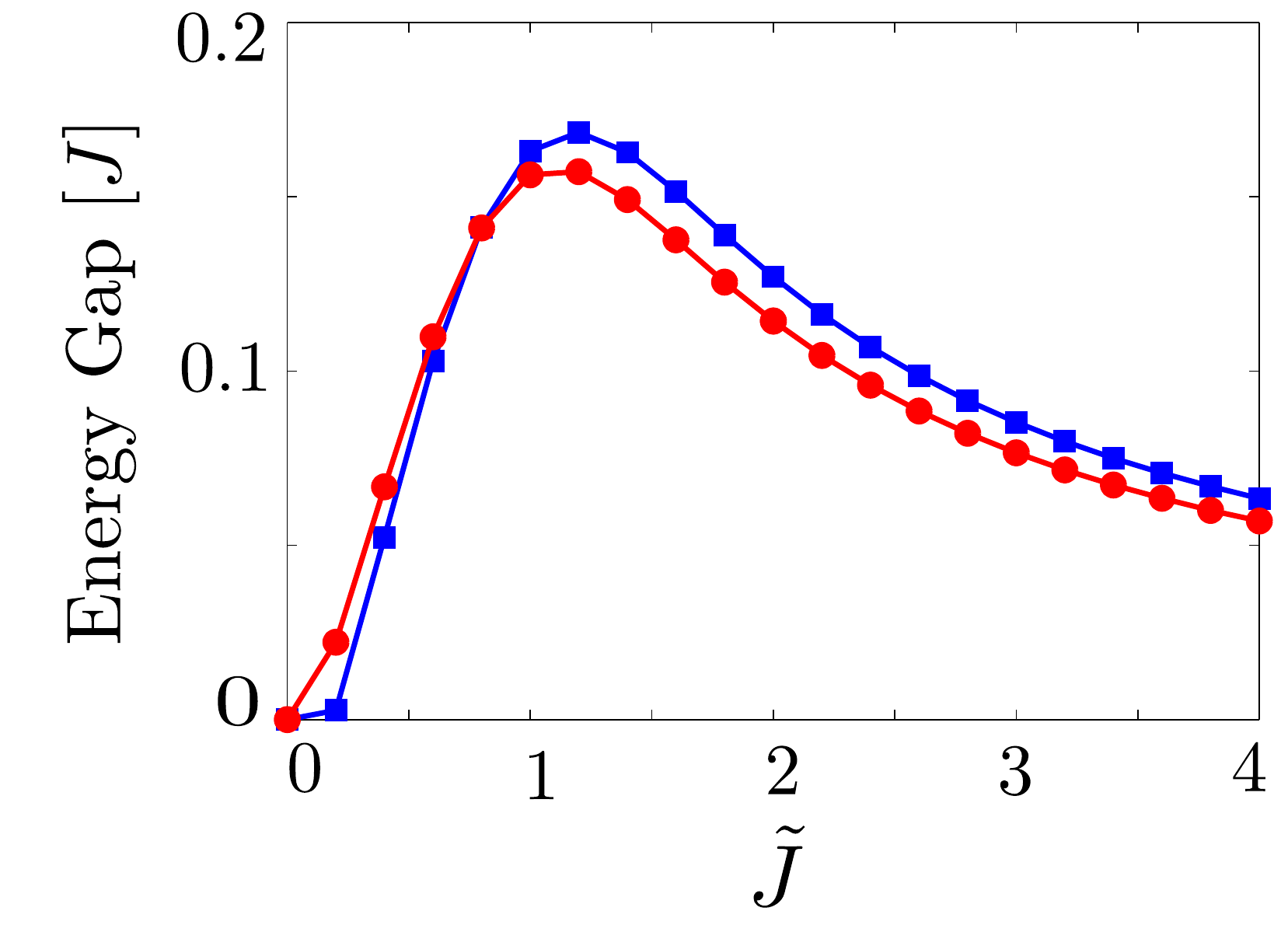}
		\caption{The minimal energy gap between the even parity state and the first excited state. {\bf Red-circles:} Ideal chain parameters: $V_e=J, N=8, \Delta=J, \mu=0.$ {\bf Blue-squares}:  Including the effects of experimental errors listed in the text. $V_e=J, N=8, \Delta=1.2J, \mu_r \in [-0.1J,0.1J]$, single-site addressability to $10\%$ accuracy and $5\%$ lagtime between lasers. }
		\label{fig:gap}
	\end{center}
	\end{figure}

To quantify the effect of these errors, we consider the size of the energy gap $\Delta_1$ between the even parity ground state and the next excited state. The magnitude of this gap is controlled by the coupling $\tilde{J}$ to the external site. In Fig.~\ref{fig:gap} we show $\Delta_1$ as a function of the coupling $\tilde{J}$ for both the ideal chain, and a non-ideal chain, including the effect of addressing errors.  From this figure we see that the errors have a very small effect on the size of the energy gap, or on the value of the ideal hopping parameter $\tilde{J}$, indicating that our protocol is robust against the class of experimental errors arising due to a non-ideal wire, and imperfect site addressing. Note that the results shown in Fig.~\ref{fig:gap} provides a lower bound on the size of the gap $\Delta_1$, which can be further increases by introducing additional asymmetry as mentioned in Section~\ref{sec:basic_idea}. The gap $\Delta_1$ sets the adiabatic time scale of the mapping protocol. For a typical value of hopping, $J\sim 500$Hz, this results in a time scale of tens of milliseconds.

	\label{sec:errors}

\subsection{Error Check}
\label{sec:error_check}
In the previous sections we have shown that while the mapping  protocol is immune to experimental errors associated with the implementation of the mapping  Hamiltonian, it is essential that both condition 1 (Eq.~\eqref{eq:energy_condition1}), and condition 2  (Eq.~\eqref{eq:energy_condition2}) are satisfied. Violating these conditions has two consequences:  most crucially, the logical states will no longer be mapped to the desired final states (as given in Eq.~\eqref{eq:desired_evolution}). In addition, the final states will be degenerate and thus the adiabatic evolution will not be reversible. 
Here, we propose methods to detect if one of these conditions has been violated. These methods will leave the logical state unaffected, ensuring that it will be done in a Quantum Non-Demolition (QND) way, and can be done on each qubit simultaneously. 

\subsubsection{Violation of Condition 1}
First, we consider the effect of violating condition 1. In this case, performing the mapping  protocol will leave the chain in an excited state, rather than the desired ground state of the Kitaev chain (see Eq.~\eqref{eq:violate_condition1}). These excited states, while degenerate, all have fixed (even) parity, compared with the odd-parity ground state. Thus, a parity measurement can be performed on the chain to distinguish between the two. If the chain is found to have even parity, it is clear an error has occurred. If the chain is found in the correct ground state, the chains must then be re-initialised in the ground state. In this procedure, the atoms on the external sites are not affected, thus the encoded information is preserved. 

\subsubsection{Violation of Condition 2}
Secondly, we consider the effect of violating condition 2. In this case performing the mapping  results in no particles occupying the external sites (see Eq.~\eqref{eq:violate_condition2}). Here we propose a protocol to detect this error by verifying if a particle is present on one of the two external sites. This protocol will not distinguish where the particle is located, thus ensuring that the protocol remains quantum non-demolition.

\begin{figure}[t!]
	\begin{center}
		\includegraphics[width=0.4\textwidth]{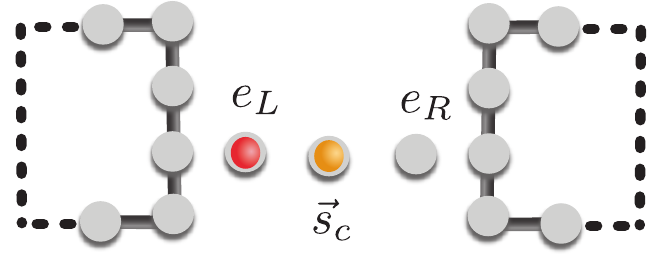}
		\caption{The geometric setup for one qubit for the proposed error check. The control qubit (shown in yellow) will be remain in the ground state if there is one particle (shown in red) on either of the external sites. If, due to an error in the mapping, no particles are present, the control qubit will be excited to an additional level which can be directly probed, as described in Section~\ref{sec:error_check} of the text.}
		\label{fig:error_check}
	\end{center}
	\end{figure}

This protocol relies on single site addressing and requires that we can excite atoms to an excited state. We require a single site $\vec{s}_c$ located between the $L$ and $R$ chain defining the qubit (see Fig.~\ref{fig:error_check}). This site hosts one fermion initialised in the ground state $|c_g\ra$ which can also be excited to an excited state $|c_e\ra$. 
Additionally, this protocol requires atoms which can be excited into a Rydberg state. Once excited into a Rydberg state, an atom interacts strongly with those within the Rybderg blockade radius, inhibiting the excitation of a second atom to the Ryberg state. Here we use this interaction to carry out an error check on each qubit, using a scheme adapted from that developed by Mueller et al. \cite{Muller09}.

The first step is to carry out a pulse sequence on the external sites $e_L,e_R$ and the site $\vec{s}_c$ . This pulse sequence is given in detail in Appendix \ref{sec:appendix_error}, and will transfer the control atom to the state

	\be
	|c_g\ra \rightarrow    |c_g\ra +(-1)^{n+1} |c_e\ra
	\ee
	
where $n$ is the number of particles on the external sites of one qubit, (i.e. $n=\la a^\dagger_{e,L}a_{e,L}+a^\dagger_{e,R}a_{e,R}\ra$). 

Second, a $\pi/2$-pulse is applied to the control site, such that the atom on site $\vec{s}_c$ will be in the $|c_g\ra$ state for an odd number of particles ($n=1$), or the $|c_e\ra$ state for an even number of particles ($n=0,2$). 

The final step is to detect the state of the control atom which will indicate if an error has occurred, while leaving the qubit state unaffected. \FloatBarrier
%========================================================================================
%========================================================================================
\section{Implementation of missing quantum gates: Controlled-Z and $\pi/8$ gate}
\label{sec:applications}
%======================================================================InterfaceImages/==================

In the previous section we introduced a mapping which coupled the topologically protected space to a conventional qubit system, and showed that it is robust against the class of experimental errors associated with a non-ideal wire and imperfect site/link addressing. Once the topological qubits have been mapped to conventional qubits, stored as the presence or absence of an atom on a single site, there are several standard techniques which can be used to manipulate them for quantum computation \cite{Briegel00}. In particular, a phase gate can be performed by exciting the atom to an excited state (with energy offset) until the time evolution ensures the desired phase. Additionally, there are several proposals for entangling gates, including collisional gates \cite{Jaksch99}, and using the long-range Rydberg interaction \cite{Jaksch00, Muller09, Brion07}. Here we consider the use of Rydberg gates for a possible implementation of a controlled-Z, as experimental setups able to both implement the Kitaev wire and carry out these gates are already developed \cite{Schausz12}. Together with the gates available from the braiding protocol, the phase gate and the controlled-Z gate are sufficient to give a complete gate set. 

\begin{figure}[t!]
	\begin{center}
		\includegraphics[width=0.45\textwidth]{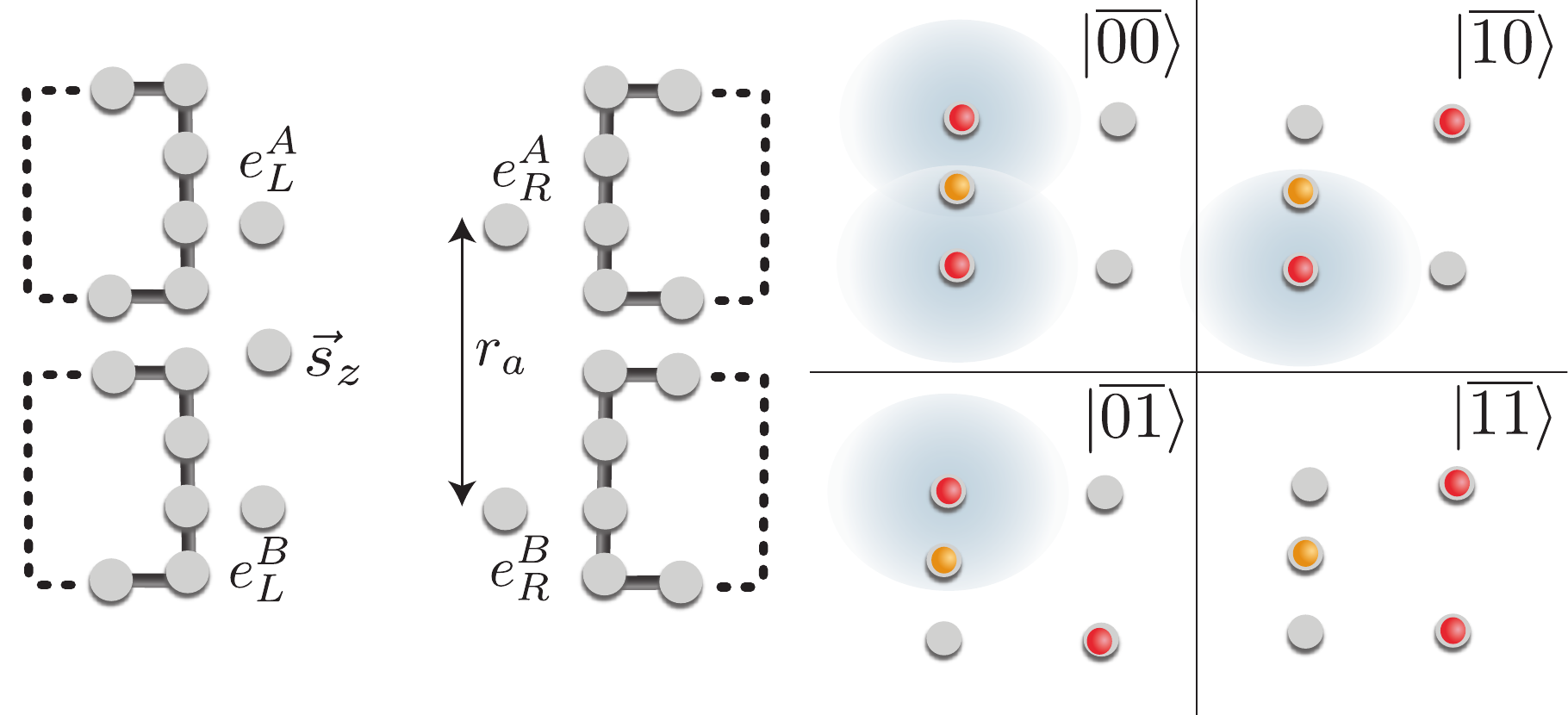}
		\caption{Schematic for the implementation of a controlled-Z gate. The external sites are labelled $e_{j}^\alpha$ where $j=L/R$ labels the left and right wire of a single qubit and $\alpha = A/B$ labels the qubit. The external sites of two atoms are separated by a distance $r_a$. The control site between qubit $A$ and $B$ is labelled $s_c^{AB}$ (shown in yellow) and is at an equal distance between the two. A controlled-Z gate can be implemented between qubit $A$ and $B$ using the physics of the Rydberg blockade and the additional control qubit \cite{Brion07}.}
		\label{fig:control_gate}
	\end{center}
	\end{figure}

The implementation of a controlled-Z gate has been discussed by Brion et al. \cite{Brion07}. In this particular setup, implementing this gate requires an additional atom positioned on site $\vec{s}_z$ between the two qubits $A$ and $B$, separated by a distance $r_a$, as shown in Fig.~(\ref{fig:control_gate}).
The particular scheme is given in more detail in Appendix~\ref{sec:appendix_gate} and is summarised here in brief.    
The first step to implement this gate is to excite any atoms on the sites $e_L^{\alpha}$, (here $\alpha= A,B$) into the Rydberg states. This excitation is done with a laser with Rabi frequency $\Omega_1 \gg V_{d}(r_a)$, for $V_{d}(r)$ the strength of the Rydberg interaction at distance $r$. This ensures that these two atoms do not interact with each other. 
The second step is to use a second laser, with Rabi frequency $\Omega_2 \ll V_{dd}(r_a/2)$ to excite the atom on the site $\vec{s}_z$ to the Rydberg state. 
Here, there are two possibilities. If the two qubit state was initially $|\overline{00}\ra, |\overline{01}\ra, |\overline{10}\ra$, due to the arrangement of atoms on the sites $e_{L}^\alpha$, the first step will cause a blockade on the atom on site $\vec{s}_z$.  However, if the two qubit state was initially $|\overline{11}\ra$, then there will be no blockade, and the atom on site $\vec{s}_z$ will be successfully excited to the Ryberg state. Finally, the atom on site $\vec{s}_z$ is brought back down to the ground state with a phase shift of $\pi$, realising a controlled-Z gate on the two qubits.

We assume in this protocol that the lasers can be focussed such that the external sites can be excited to the Rydberg states while leaving the chain in the ground state. Exciting an atom of the chain to the Rydberg state would cause a blockade on the control qubit regardless of the qubit state, rendering the gate ineffective. This assumption, however, is not crucial, as one can put extra (empty) sites between the sites $e_{R/L}$ and the chain in order to ensure that these sites are adequately separated.  % (see Fig.~\ref{fig:control_gate}). 

%========================================================================================
%========================================================================================
\section{Outlook}

In conclusion, we have presented a complete toolbox for quantum computation in a system of cold atoms stored in optical lattices. Our model takes a hybrid approach, where elements of topological quantum computing are combined with conventional quantum gates in an atomic setup. The topological elements in this hybrid model include the storage of qubits in the Majorana edge modes of a set of Kitaev wires, and the realisation of topologically protected gates via a protocol for braiding Majorana fermions. In addition, we have described a protocol to read and write the topological quantum memory which acts to map the topological Majorana qubits to conventional atomic qubits defined by the presence of single atoms. This provides not only a way to prepare and measure qubits by standard atomic and quantum optic techniques, but also allows for missing gates to be replaced by the (non-topological, i.e. unprotected) entangling gates with conventional atomic qubits, e.g. as collisional gates or Rydberg gates \cite{Briegel00, Jaksch99, Jaksch00}. 

We do not see the present hybrid quantum computing model with atoms to be in direct competition with existing quantum computing proposals and realizations with cold atoms and ions, and their remarkable achievements in laboratory implementations. In an ion trap quantum computer, for example, long lived quantum memory is achieved by selecting physical qubits, which are insensitive from the outset to perturbations, e.g.~qubits encoded in {\em clock states} or {\em decoherence free subspaces} \cite{Lidar98, Blatt12}. In addition high fidelity quantum gates are realized as a combination of high precision control of external fields, and designing gates, which are immune to the most important imperfections. In contrast, in the present atomic hybrid scenario the energy gaps underlying the error protection are typically small in comparison with realistic errors in atomic setups. Thus, the present model system should be seen more as a {\em playground}  to test the basic principles and error protection of topological quantum computing in a controlled environment. Equally important, we provide realistic atomic tools for demonstrating non-Abelian statistics of Majoranas, for example in an interferometer setup, including preparation and readout.

\section*{Acknowledgments}  
We thank G. Brennen, T. Osborne, A. Carmele and A. Glaetzle for useful discussions and M. Rider for useful comments on the manuscript. This work was supported by the SFB FoQus, the ERC Synergy Grant UQUAM and SIQS.  C.~L. is partially supported by NSERC. 
\begin{appendix}
\section{Braiding}
\label{sec:append_braiding}
Using the notation defined in Fig.~(\ref{fig:define_qubits_braiding}), the braiding operations result in the following unitary operations (with the basis $(|\overline{00}\ra, |\overline{01}\ra, |\overline{10}\ra,|\overline{11}\ra)$)
\begin{eqnarray}
U_{12} = (1-\hat{\gamma}_1\hat{\gamma}_2)/\sqrt{2} &=& \left(\begin{array}{*{20}{c}}1&0\\0&i \end{array}\right)  \otimes\mathbb{1}_2\nonumber \\ 
U_{13} = (1-\hat{\gamma}_1\hat{\gamma}_3)/\sqrt{2} &=&  \mathbb{1}_4+ \left(\begin{array}{*{20}{c}}0&-1\\1&0 \end{array}\right)  \otimes\mathbb{1}_2
\end{eqnarray}
where $\mathbb{1}_n$ is the identity matrix in $n$ dimensions.

\section{Interface Error Check}
\label{sec:appendix_error}
In this Appendix we expand on the protocol for an error check, as introduced in Section~\ref{sec:error_check}. The protocol will determine the number of particles occupying the two external sites $e_L,e_R$ associated with one qubit. For simplicity we use $|\Omega\ra\equiv |\Omega_e\ra_L\otimes|\Omega_e\ra_R, |L\ra\equiv a_{e,L}^\dagger|\Omega_e\ra_L\otimes|\Omega_e\ra_R$ and $|R\ra\equiv|\Omega_e\ra_L\otimes a_{e,R}^\dagger|\Omega_e\ra_R$. %and $|L,R\ra\equiv a_{e,L}^\dagger|\Omega_e\ra_L\otimes a_{e,R}^\dagger|\Omega_e\ra_R$

The pulse sequence is as follows:
\begin{enumerate}
\item Perform a $3\pi/2$ pulse on the control atom and a $\pi/2$ pulse on the atoms on the external sites 
\begin{eqnarray}
|c_g\rangle|\Omega\rangle &\rightarrow &( |c_g\rangle-|c_e\rangle)|\Omega\rangle/\sqrt{2} \nonumber \\
|c_g\rangle|L\rangle &\rightarrow &( |c_g\rangle-|c_e\rangle)|L_{+}\rangle/\sqrt{2} \nonumber \\
|c_g\rangle|R\rangle &\rightarrow &( |c_g\rangle-|c_e\rangle)|R_{+}\rangle/\sqrt{2} 
\end{eqnarray}
where $L(R)_+$ denotes the superposition between the ground and excited states. 
\item Perform the protocol as outlined in \cite{Muller09}. This protocol results in the acquisition of the phase $\phi$ if there is one atom on either site $e_L$ or $e_R$. As shown in  \cite{Muller09} this phase can be chosen to satisfy $\phi =\pi$, thus obtaining
\begin{eqnarray}
 ( |c_g\rangle-|c_e\rangle)|\Omega\rangle/\sqrt{2} &\rightarrow &  ( |c_g\rangle-|c_e\rangle)|\Omega\rangle/\sqrt{2} \nonumber \\
( |c_g\rangle-|c_e\rangle)|L_{+}\rangle/\sqrt{2} &\rightarrow &( |c_g\rangle-e^{i\pi}|c_e\rangle)|L_{+}\rangle/\sqrt{2} \nonumber \\
( |c_g\rangle-|c_e\rangle)|R_{+}\rangle/\sqrt{2} &\rightarrow &( |c_g\rangle-e^{i\pi}|c_e\rangle)|R_{+}\rangle/\sqrt{2}\nonumber \\
\end{eqnarray}
\item Perform a global $\pi/2$ pulse on all atoms. 
\begin{eqnarray}
( |c_g\rangle-|c_e\rangle)|\Omega\rangle/\sqrt{2}  &\rightarrow & |c_e\rangle|\Omega\rangle  \nonumber \\
( |c_g\rangle-e^{i\pi}|c_e\rangle)|L_{+}\rangle/\sqrt{2}  &\rightarrow &|c_g\rangle|L\rangle \nonumber \\
( |c_g\rangle-e^{i\pi}|c_e\rangle)|R_{+}\rangle/\sqrt{2} &\rightarrow& |c_g\rangle|R\rangle 
 \end{eqnarray}
\end{enumerate}
Therefore, by measuring the control qubit in the excited state one can read out if there has been an error in the protocol resulting in no extracted particles. Because the effect of a particle on the left  $e_L$ or right, $e_R$ external site is equivalent, this protocol is QND, giving no information on the qubit information.

\section{Entangling Gate}
\label{sec:appendix_gate}
\begin{figure}[t!]
\begin{center}
\includegraphics[width=0.4\textwidth]{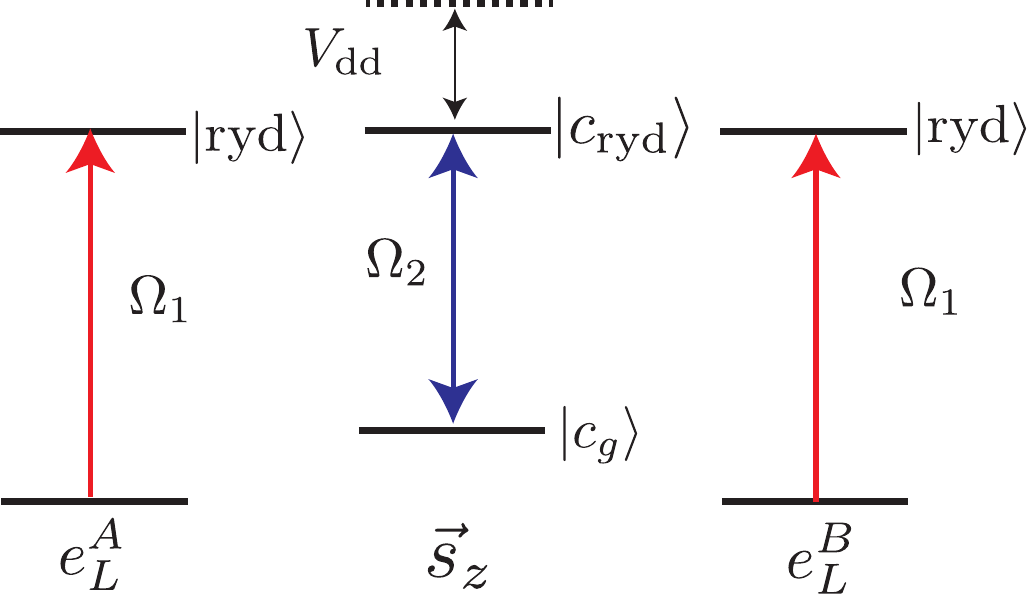}
\caption{ Pulse sequence for implementing a Control-Z gate. If particles are present on the sites $e_L^A$ or $e_L^B$ they will be excited to the Rydberg state $|{\rm ryd}\ra$ via a laser with Rabi frequency $\Omega_1 \gg V_{dd}$ (shown in red), causing a Rydberg blockade $V_{dd}$ on the atom on site $\vec{s}_z$. A laser of Rabi frequency $\Omega_2 \ll V_{dd}$ excites the atom on site $\vec{s}_z$ from the ground state to the Rydberg state. Due to the blockade, this process will only be successful if there were initially no particles on sites $e_L^A$ or $e_L^B$. }
\label{fig:control_gate_pulses}
\end{center}
\end{figure}

 In this Appendix we give the detailed pulse sequence required to implement the controlled-Z gate, as outlined in Section~\ref{sec:applications}. The pulse scheme is shown schematically in Fig.~\ref{fig:control_gate_pulses}. To describe the sequence in detail, we follow the evolution of the four possible two-qubit logic states. Using the same notation as above, $ |L\ra\equiv a_{e,L}^\dagger|\Omega_e\ra_L\otimes|\Omega_e\ra_R$ and $|R\ra\equiv|\Omega_e\ra_L\otimes a_{e,R}^\dagger|\Omega_e\ra_R$ the four logical states are
\begin{eqnarray}
|\overline{00}\ra \xleftrightarrow{\text{mapping}} |L\ra^A|L\ra^B \nonumber \\
|\overline{01}\ra \xleftrightarrow{\text{mapping}} |L\ra^A|R\ra^B \nonumber \\
|\overline{10}\ra \xleftrightarrow{\text{mapping}} |R\ra^A|L\ra^B \nonumber \\
|\overline{11}\ra \xleftrightarrow{\text{mapping}} |R\ra^A|R\ra^B 
 \end{eqnarray}
 where $A,B$ labels the two qubits. 
The protocol is as follows
\begin{enumerate}
\item A pulse with Rabi frequency $\Omega_1$ on the sites $e_L^{A}$ and $e_L^{B}$ excites any atom present on these sites to the Rydberg level $|{\rm ryd}\ra$. We assume $\Omega_1 \ll V_{dd}(r_a)$, where $r_a$ is the distance between sites $e^A_L$ and $e_L^B$ (see Fig.~\ref{fig:error_check}). This ensures that these atoms do not interact. The result is
\begin{eqnarray}
|\overline{00}\ra &\rightarrow&|{\rm ryd}\ra^A|{\rm ryd}\ra^B \nonumber \\
|\overline{01}\ra &\rightarrow& |{\rm ryd}\ra^A|R\ra^B \nonumber \\
|\overline{10}\ra &\rightarrow&|R\ra^A|{\rm ryd}\ra^B \nonumber \\
|\overline{11}\ra &\rightarrow &|R\ra^A|R\ra^B 
 \end{eqnarray}
This is shown by the red pulses in Fig.~(\ref{fig:control_gate_pulses}).

\item A second pulse with Rabi frequency $\Omega_2\ll V_{dd}(r_a/2)$ on the site $\vec{s}_z$ to excite the atom on this site to the Rybderg state . In the case of logical states $|\overline{00}\ra,|\overline{01}\ra ,|\overline{10}\ra $ the excitation of this atom is blocked by the Rydberg blockade induced by the atoms on sites $e_L^\alpha$ already occupying the Rydberg state. However in the case of the logical state $|\overline{11}\ra$ there is no blockade, and the atom on site $\vec{s}_z$ is successfully excited into the Rydberg state. 

\item A pulse on site $\vec{s}_z$ to de-excite the atom back to the ground state, with a phase shift of $\pi$. Because the atom on site $\vec{s}_z$ is only in the Rydberg state if the initial qubit state was $|\overline{11}\ra$, this state alone will pick up this phase shift.

\item A pulse to bring all atoms on sites $e_L^\alpha$ back to the ground state. 

\end{enumerate}
This protocol results in a Controlled-Z gate acting on the logical subspace
\begin{eqnarray}
|\overline{00}\ra &\rightarrow & |\overline{00}\ra \nonumber \\
|\overline{01}\ra  &\rightarrow & |\overline{01}\ra \nonumber \\
|\overline{10}\ra  &\rightarrow & |\overline{10}\ra \nonumber \\
|\overline{11}\ra  &\rightarrow & -|\overline{11}\ra .
 \end{eqnarray}

\end{appendix}

\bibliographystyle{apsrev}
\bibliography{CLRefs}
\end{document}